\documentclass[12pt,a4,sans]{article}
\pdfoutput=1
\usepackage[latin1]{inputenc}
\usepackage{amssymb}
\usepackage{amsmath}
\usepackage{amsthm}
\usepackage{amsfonts}
\usepackage[pdftex]{graphicx}
\usepackage{geometry}
\usepackage{bbold}
\usepackage{braket}
\usepackage{enumerate}
\usepackage[usenames,dvipsnames]{xcolor}
\usepackage{setspace}
\usepackage[backgroundcolor=white,linecolor=black]{todonotes}
\usepackage[in]{fullpage}
\usepackage{hyperref}
\usepackage{array}
\usepackage{cancel}
\usepackage{wrapfig}
\usepackage[font=small,labelfont=bf]{caption}
\usepackage{anysize}

\numberwithin{equation}{section}

\marginsize{2.4cm}{2.4cm}{2cm}{2cm}

\hypersetup{
colorlinks=true,
linktoc=page,
citecolor=PineGreen,
linkcolor=BlueViolet,
urlcolor=MidnightBlue} 

\makeatletter 
\renewcommand{\maketitle} 
 { \begingroup \begin{center} \large {\bf \@title}
 	\vskip 5pt \large \@author \\ \vskip 5pt \@date \end{center}
   \vskip 5pt \endgroup \setcounter{footnote}{0} }
\makeatother 

\newcommand{\comments}[1]{}
\newcommand{\la}{\langle}
\newcommand{\ra}{\rangle}
\newcommand{\tl}{\widetilde\lambda}

\newcommand{\A}{\mathcal{A}}
\newcommand{\F}{\mathcal{F}}
\newcommand{\T}{\mathcal{T}}

\newcommand{\N}{\mathcal{N}}

\newcommand{\Tr}{\text{Tr}}
\newcommand{\MHVb}{\overline{\text{MHV}}}
\renewcommand{\b}[1]{\braket{#1}}

\renewcommand{\O}{\mathcal{O}}

\newcommand{\be}{\begin{equation}}
\newcommand{\ee}{\end{equation}}

\def\beqa{\begin{eqnarray}}
\def\eeqa{\end{eqnarray}}
\def\beq{\begin{equation}}
\def\eeq{\end{equation}}

\def\Tr{{\rm Tr}}
\def\one{\mbox{1 \kern-.59em {\rm l}}}

  \def\cC{{\cal C}}
  \def\cF{{\cal F}}
  
  \def\cL{{\cal L}}
 \def\cN{{\cal N}} \def\cO{{\cal O}}
  
 \def\cT{{\cal T}}

\def\uno{\mbox{1 \kern-.59em {\rm l}}}

\def\lan{\langle}
\def\ran{\rangle}

\def\one{1\!\!1\,\,}

\def\bcomment#1{}

\def\eps{\epsilon}

\def\Li{{\rm Li}_2}

\usepackage{slashed}
\usepackage{caption}

\usepackage[nottoc,notlot,notlof]{tocbibind}
\usepackage[nosort]{cite}
\usepackage{pstricks}         
\usepackage{color}
\usepackage{bbm}
\usepackage{parskip}

\long\def\symbolfootnote[#1]#2{\begingroup%
\def\thefootnote{\fnsymbol{footnote}}\footnote[#1]{#2}\endgroup}

\begin{document}

\begin{flushright}
QMUL-PH-14-01
\end{flushright}

\vspace{20pt}

\begin{center}

{\Large \bf On  super  form factors of half-BPS operators  }
\\
\vspace{0.4cm}
{\Large \bf  in  $\cN=4$ super Yang-Mills  }

\vspace{45pt}

{\mbox {\bf  Brenda Penante, Bill Spence,  Gabriele Travaglini and   Congkao Wen}}%
\symbolfootnote[4]{
{\tt  \{ \tt \!\!\!b.penante, w.j.spence, g.travaglini, c.wen\}@qmul.ac.uk}
}

\vspace{0.5cm}

\begin{center}
{\small \em

Centre for Research in String Theory\\
School of Physics and Astronomy\\
Queen Mary University of London\\
Mile End Road, London E1 4NS, UK
}
\end{center}


\vspace{40pt}

{\bf Abstract}
\end{center}

\vspace{0.3cm}

\noindent
We compute form factors of  half-BPS operators in $\cN=4$ super Yang-Mills dual to massive Kaluza-Klein modes in supergravity.  These are appropriate supersymmetrisations $\cT_k$ of the scalar operators $\Tr \, (\phi^k)$ for any $k$, which for $k\!=\!2$ give the chiral part of the stress-tensor multiplet operator.   
Using harmonic superspace, we derive simple Ward identities for these form factors, which we then compute perturbatively at tree level and one loop. 
We propose a novel on-shell recursion relation which links form factors with different numbers of fields. Using this,
we conjecture a general formula for the $n$-point MHV form factors of $\cT_k$ for arbitrary $k$ and $n$.
Finally, we use supersymmetric generalised unitarity to derive compact expressions for all one-loop MHV form 
factors of $\cT_k$ in terms of one-loop triangles and finite two-mass easy box functions.

\setcounter{page}{0}
\thispagestyle{empty}
\newpage


\setcounter{tocdepth}{4}
\hrule height 0.75pt
\tableofcontents
\vspace{0.8cm}
\hrule height 0.75pt
\vspace{1cm}

\setcounter{tocdepth}{2}


\pagenumbering{arabic}

\section{Introduction}

Recently, there has been a resurgence of interest in the study of form factors in $\cN\!=\!4$ super Yang-Mills (SYM). One reason behind this is that form factors interpolate between fully on-shell quantities, i.e.~scattering amplitudes, and correlation functions, which are  off shell. Indeed, form factors are obtained   by taking a gauge-invariant, local operator $\cO (x)$ in the theory, applying it to the vacuum $|0\rangle$, and considering the overlap with a multi-particle state $\langle 1, \ldots , n |$. In momentum space, the quantity we consider is
\beq
F(1, \ldots, n;q) \ := \   \int\!d^4x \, e^{- i q x} \  \langle 1, \ldots , n |\cO (x)  |0\rangle  \ = \ \delta^{(4)} \big(q - \sum_{i=1}^n p_i\big) 
  \langle 1, \ldots , n |\cO (0)  |0\rangle\ , 
\eeq
where the delta function is a simple consequence of translational invariance of the theory. Once we fix a certain operator, one can study how the form factor changes as we vary the state.  

In a pioneering paper \cite{vanNeerven:1985ja} almost thirty years ago, van Neerven considered the simplest form factor of the simplest half-BPS operator in $\cN=4$ SYM, namely the two-point (or Sudakov) form factor of the scalar operator $\Tr (\phi^2) $,  deriving its expression at one and two loops. More recently,  the computation of form factors at strong coupling was considered in \cite{Alday:2007he,Maldacena:2010kp}, and shortly after at weak coupling in a number of papers  in $\cN=4$ SYM \cite{Brandhuber:2010ad,Bork:2010wf,Brandhuber:2011tv,Bork:2011cj,Broedel:2012rc,Henn:2011by,Gehrmann:2011xn,Brandhuber:2012vm,Bork:2012tt,Engelund:2012re,Boels:2012ew} and also in the ABJM theory \cite{Brandhuber:2013gda,Young:2013hda,Bianchi:2013pfa}.   In particular, in \cite{Brandhuber:2010ad} it was pointed out that on-shell methods can successfully be applied to the computation of such quantities, and the expression for the infinite sequence of MHV form factors of the simplest dimension-two, scalar half-BPS operators was computed. Perhaps unsurprisingly, this computation revealed the remarkable simplicity of this quantity -- for instance, the form factor of two scalars and $n-2$ positive-helicity gluons is very reminiscent of the  Parke-Taylor MHV amplitude \cite{ParkeTaylor:1986,Mangano:1987xk},  
\beq
\lan  g^+(p_1) \cdots  \phi_{12} (p_i) \cdots    
 \phi_{12} (p_j) \cdots g^+ (p_n) | \cO (0) | 0 \ran \ = \ {\lan ij\ran^2 \over \lan 12\ran \cdots \lan n1\ran }
\ , 
\eeq
  where%
  \footnote{More generally,  the scalar, bilinear half-BPS   operators in $\cN=4$ SYM  can be defined as 
  $\cO_{ABCD} := 
{\rm Tr} (\phi_{AB} \phi_{CD})  \, - \,  (1/12) \,  \eps_{ABCD} {\rm Tr} ( \bar\phi^{LM} \phi_{LM} )$, 
where $\bar \phi^{AB} := (1/2) \eps^{ABCD} \phi_{CD}$.
These operators  belong to the  $\mathbf{20}^\prime$ representation of the  $SU(4)$  R-symmetry group.}   
$\cO := \Tr [ (\phi_{12})^2]$.    These form factors maintain this simplicity also at one loop -- they are proportional to their tree-level expression, multiplied by a sum of one-mass triangles and two-mass easy box functions. Other common features between form factors and amplitudes include the presence of a version of colour-kinematics duality \cite{Boels:2012ew} similar to that of BCJ \cite{Bern:2008qj}, and the possibility of computing form factors at strong coupling using Y-systems \cite{Maldacena:2010kp,Gao:2013dza} which extend those of the  amplitudes  \cite{Alday:2010vh}.
 A second motivation to study form factors is therefore to explore to what extent their  simplicity is preserved as we vary the choice of the operator and of the external state. 

There are interesting distinctive features of   form factors as compared to scattering amplitudes. One of them is the presence of non-planar integral topologies in their perturbative expansion. Indeed, the presence of a colour-singlet   operator introduces an element of non-planarity in the computation even when we consider   external states that are colour ordered, as  is usual in scattering amplitudes. Specifically, the external leg carrying  the momentum of the operator does not participate in the colour ordering, and hence non-planar integrals are expected to appear  at loop level. Even the simple  two-loop Sudakov form factor of  \cite{vanNeerven:1985ja} is expressed in terms of a planar as well as a non-planar two-loop triangle integral. Partly because of this nonplanarity, the novel and powerful on-shell methods of \cite{ArkaniHamed:2012nw} have not yet been extended to form factors.

One may wonder if higher-loop corrections can spoil the simple structures observed at tree level and one loop. There is a number of examples which indicate that, fortunately, this is not the case. For instance, in \cite{Gehrmann:2011xn} the three-loop corrections to the Sudakov form factor were computed and found to be given by a maximally transcendental expression, which furthermore happens to be equal to the leading transcendentality part of the quark and gluon form factors in QCD. 
A remarkably compact result was found in 
\cite{Brandhuber:2012vm} for the three-point form factor of half-BPS bilinear operators in $\cN=4$ SYM at two loops. This result is expressed in terms of planar and non-planar double-box integrals, as well as two-loop triangle functions. Exponentiation of the infrared divergences leads one to define a finite remainder function very much in the same spirit of the BDS remainder function \cite{Bern:2008ap,Drummond:2008aq}. Using the concept of the symbol of a transcendental function  \cite{Goncharov:2010jf} as well as various physical constraints, it was found that the form factor remainder is given by a remarkably simple, two-line expression written in terms of classical polylogarithms only. Moreover, the remainder function  was found to be closely related to the analytic expression of the six-point remainder at two-loops found in \cite{Goncharov:2010jf}. 

Similarly to the miraculous simplifications which occur in going from the result of an explicit calculation \cite{DelDuca:2010zg} to the expression of  \cite{Goncharov:2010jf}, the (complicated) two-loop planar and non-planar functions found in \cite{Brandhuber:2012vm} combine into a maximally transcendental, compact result. Surprising agreement was furthermore found between this form factor and the maximally transcendental part of certain very different quantities, namely  the Higgs plus three-gluon amplitudes in QCD computed in \cite{Gehrmann:2011aa}. A hint of a possible connection between such unrelated quantities (and a further reason to study half-BPS form factors in $\cN=4$ SYM) is that the top component of the stress-tensor multiplet operator (of which $\Tr \, (\phi^2)$ is the lowest component) is the on-shell Lagrangian of the theory, which contains the term  $\Tr \, F_{\rm SD}^2$, where $F_{\rm SD}$ is the self-dual part of the field strength. In turn, it is known that Higgs plus multi-gluon amplitudes in the large top mass limit can be obtained from  an effective interaction of the form $H \,\Tr \, F_{\rm SD}^2$ \cite{Wilczek:1977zn, Shifman:1978zn} (see also \cite{Dixon:2004za} for a recent discussion). Inserting this interaction once is precisely equivalent to computing the form factor of $\Tr \, F_{\rm SD}^2$. 

Incidentally, we note that form factors can be used to compute correlation functions using generalised unitarity as in \cite{Engelund:2012re}. They also appear in the intermediate sums defining total cross sections, or the event shapes considered in \cite{Belitsky:2013ofa,Belitsky:2013bja,Belitsky:2013xxa}.

So far, most of the available results are concerned with bilinear half-BPS operators.%
\footnote{With the exception of \cite{Bork:2010wf}, where form factors of operators of the form $\Tr \, (\phi^n)$ were considered with an external state containing the same number $n$ of particles as of fields in the operator. These can be thought of as a generalisation of the Sudakov form factor, for which $n=2$. } In this paper we will focus on form factors of operators of the form $\Tr \, (\phi^k)$ with an $n$-point external state, for arbitrary $k$ and $n$. These operators are dual to massive Kaluza-Klein modes of the $AdS_5 \times S^5$ compactification of 
type IIB supergravity for $k\geq 3$, and their four-point functions were studied in \cite{Arutyunov:2002fh}.
For $k=2$ these operators are part of the stress-tensor multiplet, and are dual to the massless graviton multiplet. \
In fact, there is no reason to limit our study to scalar operators, as one can supersymmetrise the scalar operators in a similar fashion as  is done in the case of the stress-tensor multiplet operator.  Thus, the  operator we consider is  
\beq
\label{tk}
\cT_k\  := \ \Tr[ (W^{++})^k]
\ , 
\eeq
where $W^{++}$  is a particular projection  
of the chiral vector multiplet superfield  $W^{AB}(x, \theta)$ of $\cN=4$ SYM,  introduced   in the next section.    
For $k=2$ this is the chiral part of  the stress-tensor multiplet operator.  
$\cT_k$  is a half-BPS operator, and its lowest component is simply  the scalar  operator $\Tr [(\phi^{++})^k]$. 

In Section \ref{sec:Ward} we review a convenient formalism to study these operators, namely  harmonic superspace  \cite{Galperin:1984av,Galperin:2001uw}.  
We will then consider form factors of the  chiral part of the operators $\cT_k$, which preserve half of the supersymmetries off shell \cite{Eden:2011yp, Eden:2011ku}.  External states will be described  naturally  with the  supersymmetric formalism of   Nair \cite{Nair:1988bq}. One can then write down very simple Ward identities, similar to those considered in \cite{Brandhuber:2011tv} for the case of the stress-tensor multiplet operator, which we can then solve finding constraints on the  expressions for the form factors.

In Section \ref{sec:susyff} we consider the simplest supersymmetric form factors, namely those of $\cT_3$. Using  BCFW recursion relations \cite{BCF:RecursionRelations,Britto:2005} (in the supersymmetric version of  \cite{ArkaniCachazo:2008SimplestQFT,Brandhuber:2008pf}) we will find a compact expression for the  $n$-point form factor of this operator. Interestingly, the standard recursion relation with adjacent shifts contains a boundary term, hence we are led to use a recursion relation with next-to-adjacent shifts.   

The presence of boundary terms in the adjacent-shift recursion relations for the form factor of $\cT_3$ motivates us in Section \ref{sec:adjBCFWshift} to study their structure for the case of the form factor of $\cT_k$ for general $k$. This will lead us to propose a new supersymmetric recursion relation for the MHV form factors of $\cT_k$, which involves form factors with different operators, namely $\cT_k$ and $\cT_{k-1}$. We also look at a simple generalisation of this recursion to the case of NMHV form factors. Based on some experimentation for lower values of $k$, we propose a general solution for all $n$-point MHV form factors of $\cT_k$ for arbitrary $k$ and $n$. We also check that our proposed solution satisfies the required cyclic symmetry. 

Section \ref{sec:MHVrules} briefly shows that MHV diagrams \cite{Cachazo:2004kj} can be extended  to compute form factors of the half-BPS operators considered in this paper, as a simple extension of the work of  \cite{Brandhuber:2011tv} where MHV rules for the stress-tensor multiplet operator were found. 
We present two examples in detail, namely the calculation of a four-point NMHV form factor using bosonic as well as supersymmetric MHV rules. 

In Section \ref{sec:1loopTk} we move on to one-loop level. We begin by deriving the universal form of the infrared-divergent part of generic form factors in $\N=4$ SYM.  This is determined by a single two-particle diagram where a four-point amplitude sits on one side of the cut.  We then compute the three-point form factor of $\cT_3$ at one loop, and then extend this result to $n$ points using supersymmetric quadruple cuts \cite{Drummond:2008bq}. 
Finally, we present the expression for the infinite sequence of $n$-point MHV form factors of $\cT_k$ for arbitrary $k$  and $n$.


\section{Super form factors  of $\cT_k$ and Ward identities}
\label{sec:Ward}

In this section we will study the  supersymmetric form factors of the operators $\cT_k$ introduced in \eqref{tk}, 
which generalise  those of the stress-tensor multiplet operator studied in  \cite{Brandhuber:2011tv}.


We begin our discussion  by recalling   that  the states in  the $\N=4$ multiplet can be efficiently  described using the formalism introduced by  Nair \cite{Nair:1988bq}. 
This is based on the introduction of a super-wavefunction 
\beq
\label{eq:supermultiplet}
\Phi(p,\eta) := g^+(p) + \eta_A \lambda^A(p) + {\eta_A \eta_B\over
2!} \phi^{AB}(p) + \epsilon^{ABCD} {\eta_A \eta_B \eta_C\over 3!}
\bar\lambda_D(p) + \eta_1 \eta_2 \eta_3 \eta_4 g^-(p)
\, ,
\eeq
where  $\eta_A$ is a Grassmann variable, and $A=1, \ldots , 4$ is a fundamental R-symmetry index. Here  $\big(g^+(p), \ldots , g^{-}(p)\big)$ denote the annihilation operators of the corresponding states.  In order to select a state with 
helicity $h_i$, one simply  expands the superamplitude and picks the term with $2 - 2h_i$ powers of  $\eta_i$.

The  supersymmetric operator we wish to consider is a  generalisation of the chiral part of the stress-tensor multiplet operator $\T_2$. 
It is defined as 
\begin{equation}
\label{T}
\T_k (x, \theta^{+}) := {\Tr} \big[\big(W^{++} (x, \theta^{+})\big)^k\big]\ , 
\end{equation}
where $W^{++}$ is a particular projection of the chiral vector multiplet superfield  $W^{AB}(x, \theta)$, defined as follows.%
\footnote{We follow closely the notation and conventions of 
\cite{Eden:2011yp,Eden:2011ku}, see also \cite{Brandhuber:2011tv}.} 
We introduce the harmonic projections of  the chiral superspace coordinates  $\theta^A_{\alpha}$  and supersymmetry charges $Q^{\alpha}_A$
as 
\begin{align}
\theta_{\alpha}^{\pm a} \,:=\,  \theta^A_{\alpha} u_A^{\pm a}\, \qquad  \ Q_{\pm a}^{\alpha} \,:=\, \bar{u}^A_{\pm a} Q^\alpha_A \, .
\end{align}
Here  $a\,=\,1,2$ is an $SU(2)$ index, and the
harmonic $SU(4)$  $u$ and $\bar{u}$ variables are normalised as in Section 3 of
\cite{Eden:2011yp}.  Then 
\be
W^{+a+b}\, := \, u_A^{+a} u_B^{+b}  W^{AB}   \, = \, \eps^{ab} \, W^{++} 
\ . 
\eeq
In particular, the chiral part of the stress-tensor multiplet operator is simply  
\beq
\label{cpst}
\T_2 (x, \theta^{+})\, := \, \Tr( W^{++}W^{++}) (x, \theta^{+})
\, = \, \Tr ( \phi^{++}\phi^{++})\, + \, \cdots \, + \,  {1\over 3} (\theta^{+})^4 \cL\ . 
\eeq
Note that the $(\theta^+)^0$ component is the scalar operator 
$\Tr (\phi^{++} \phi^{++})$, whereas the $(\theta^+)^4$ component is the chiral on-shell Lagrangian  denoted by $\cL$. In complete analogy to \eqref{cpst}, we have 
\beq
\T_k (x, \theta^{+}) \, = \, \Tr \big[( \phi^{++})^k\big] \, + \, \cdots 
\ . 
\eeq
Ward identities associated to supersymmetry can be used to constrain the expression of the super form factor. This was done in \cite{Brandhuber:2011tv} and we briefly review here this procedure.  
We consider a symmetry generator $s$ that annihilates the vacuum. It then follows that 
\beq
\lan 0 | [s \, ,   \Phi(1) \cdots \Phi (n)\, \cO \,] | 0  \ran  \ = \ 0  
\ ,
\eeq
or
\beq
\label{WI}
 \lan 0 |  \Phi(1) \cdots \Phi (n)\,  [s \, , \, \cO] \,| 0  \ran \, + \, \sum_{i=1}^n  \lan 0 | \,  \Phi(1) \cdots [s \, , \, \Phi(i) ] \cdots \Phi (n)\, \cO | 0  \ran \ = \ 0
\ , 
\eeq
where   $\lan 0 | \Phi(1) \cdots \Phi(n)  $ is the superstate $\lan1 \cdots n |$. In this notation, a form factor is simply $\lan 0 | \Phi(1) \cdots \Phi (n)\,\cO \, | 0  \ran$ or, more compactly,  $\lan 1\cdots n | \, \cO \, | 0  \ran$.
We are  interested in the action of the supersymmetry charges $Q_{\pm}$, which  are realised  on the half-BPS operators $\cT_k$ as  
\beq
\label{two} [Q_{-}  \, , \, \cT_k (x, \theta^{+} ) ] \ = \ 0
 \ , \qquad
 [Q_{+}  \, , \, \cT_k (x, \theta^{+} ) ] \ = \  i {\partial \over \partial \theta^{+} } \cT_k (x, \theta^{+} )  \ .
 \eeq
The first relation is a simple consequence of the fact that $\cT_k (x, \theta^{+} )$ is independent of $\theta^{-}$, while the second shows that $Q_{+}$ can be used to relate the various components in the supermultiplet described by $\cT_k (x, \theta^{+} )$.

We now introduce the object we will compute, i.e.~the (super) Fourier transform of the form factor, 
\beq
\cF_{\cT_k,n} (1, \ldots, n;  q, \gamma_{+} )
\ := \ \int\!\!d^4x\, d^4 \theta^{+} \
e^{-(iqx + i \theta_\alpha^{+a}
\gamma_{+a}^\alpha) } \, \langle  \, 1 \cdots n\,  | \cT_k(x, \theta^+)
\,  | 0   \ran\ \, .
\eeq
The Ward identities \eqref{WI} for $Q_{+}$ and $Q_{-}$ give then
\beqa \label{WIqpm}
\begin{split}
 \big( \sum_{i=1}^n \lambda_i \eta_{-,
i} \big) \ \cF_{\cT_k} (1, \ldots, n;q, \gamma_{+}) & = & 0\ ,
 \\
\big( \sum_{i=1}^n \lambda_i \eta_{+, i} \,  -  \, \gamma_{+} \big)
\ \cF_{\cT_k} (1, \ldots, n; q, \gamma_{+}) & = & 0 \, ,
\end{split}
\eeqa
where 
\beq
\eta_{\pm a, i} \, := \, \bar{u}_{\pm a}^A \eta_{A, i}\ .
\eeq
Momentum conservation follows from the Ward identity for the momentum generator, 
\beq
\big(q - \sum_{i=1}^n p_i\big) \cF_{\cT_k} (1, \ldots, n ;q, \gamma_{+})\  = \ 0
\ .
\eeq
Hence, the Ward identities require that 
\beq
\cF_{\cT_k,n} (1, \ldots, n; q, \gamma_{+}) \ \propto \    \delta^{(4)}\big(q-\sum\limits_{i=1}^n\lambda_i\tl_i\big) \delta^{(4)}\big(\gamma_+-\sum\limits_{i=1}^n\lambda_i\eta_{+,i}\big) \delta^{(4)}\big(\sum\limits_{i=1}^n\lambda_i\eta_{-,i}\big)\, . 
\eeq
It was shown in   \cite{Brandhuber:2011tv} that the supersymmetric MHV form factor of the the stress-tensor multiplet operator $\T_2$ is simply obtained by multiplying the required delta functions by  a Parke-Taylor denominator:  
\begin{equation} 
\label{eq:stresstensorFF}
\F^{\text{MHV}}_{\T_2,n}(1,\ldots,n;q,\gamma_+) =\frac{  \delta^{(4)}\big(q-\sum\limits_{i=1}^n\lambda^i\tl^i\big) \delta^{(4)}\big(\gamma_+-\sum\limits_{i=1}^n\lambda_i\eta_{+,i}\big) \delta^{(4)}\big(\sum\limits_{i=1}^n\lambda_i\eta_{-,i}\big)}{\la 12\ra\la 23\ra\cdots\la n1\ra}\ . 
\end{equation}
One of the goals of this paper is to determine the form factors of the more general operators $\T_k$ for any $k$ and for a generic number $n$ of external particles.

\section{The  super form factor $\F^{\text{MHV}}_{\cT_3,n}$}
\label{sec:susyff}

In this section we will study the  form factors of the chiral operator $\cT_3$, where $\cT_k$ is defined in \eqref{T}. In particular we will consider the form factor with the simplest helicity assignment, namely MHV,%
\footnote{Note that in general, the MHV form factor of $\cT_k$  will have  fermionic degree $8+ 2(k-2)$.}
and will show that it is given by the compact expression
\begin{equation}
\label{eq:FF3MHVsusy-simple}
\F^{\text{MHV}}_{\T_3,n}(1,\dots,n;q,\gamma_{+}) \ =\ \F^{\text{MHV}}_{\T_2,n}(1,\ldots,n;q,\gamma_{+})
\  
\Big(\sum\limits_{i < j=1}^{n} \b{i\,j}\eta_{-, i} \cdot\eta_{-, j} \Big) \ , 
\end{equation}
where we have introduced the shorthand notation  $\eta_{-, i}\cdot\eta_{-,j}\, :=\, \frac{1}{2}\, \eta_{-a, i}\eta_{-b, j}\, \epsilon^{ab}$.
Interestingly, this form factor can be written as a product of the stress-tensor MHV form factor \eqref{eq:stresstensorFF} with an additional term which compensates for the different R-charge of the operator $\T_3$. Indeed, 
it is immediate to see that, for  $\F^{\text{MHV}}_{\T_3,n}$ to be non-vanishing for an external state containing three scalars and an arbitrary number of  positive-helicity gluons, the form factor  must have a fermionic degree which exceeds that of  $\F^{\text{MHV}}_{\T_2,n}$ by two units.

We also show an equivalent  expression for the super form factor  $\F^{\text{MHV}}_{\T_3,n}$ given by the following formula, 
\begin{align}
\label{eq:FF3MHVsusy}
\begin{split}
\F^{\text{MHV}}_{\cT_3,n}(1,\dots,n;q,\gamma_+) \ =\ \F^{\text{MHV}}_{\T_2,n}(1,\ldots,n;q,\gamma_+)
\, \Big(\sum\limits_{i\leq j=1}^{n-2}(2-\delta_{ij})\dfrac{\b{n\,i}\b{j\,n-1}}{\b{n-1\,n}}\eta_{-, i}\cdot\eta_{-, j}\Big)\, .
\end{split}
\end{align}
Although \eqref{eq:FF3MHVsusy} looks slightly more complicated than \eqref{eq:FF3MHVsusy-simple}, this expression will  prove more convenient for later generalisations to higher $k$ and applications to loop computations. \\[12pt]
To prove the equivalence of \eqref{eq:FF3MHVsusy-simple} and \eqref{eq:FF3MHVsusy}, consider the  expression
\begin{equation}
\label{foll}
\sum\limits_{i < j=1}^{n} \b{i\,j}\eta_{-, i}\cdot\eta_{-,j} + \sum_{i,j=1}^n
\frac{\b{n\,i}\b{j\,n-1}}{\b{n-1\,n}} \eta_{-, i}\cdot\eta_{-, j} \, .  
\end{equation}
The second term on the right-hand side of  \eqref{foll} is in fact zero due to supermomentum conservation in the $Q_{-}$ direction. Splitting the sum over all $i,j$ in that term into the cases $i=j,\, i<j$ and $j<i$, it is straightforward to show that \eqref{eq:FF3MHVsusy-simple} and \eqref{eq:FF3MHVsusy} are equal. 

We also comment that it is straightforward to show that the expression \eqref{eq:FF3MHVsusy-simple} is cyclically invariant --  defining $V(1,2,\dots,n)= \sum_{i < j=1}^{n} \b{i\,j}\eta_{-, i} \cdot\eta_{-, j}$, due to supermomentum conservation $V(1,2,\dots,n)= V(2,3,\dots,n)=V(1,2,\dots,n-1)$, whence $V(2,3,\dots,n,1) = V(2,3,\dots,n) = V(1,2,\dots,n)$.

For the case of three external legs, the form factor $\F_{\T_3}$ is simply equal to one, or  $(\eta_{-,1})^2(\eta_{-,2})^2(\eta_{-,3})^2$  
in the supersymmetric language. Indeed, it is easy to check that  \eqref{eq:FF3MHVsusy-simple} evaluated for $n=3$  reproduces this result. 
Having  established the correctness of $\F_{\T_3}$ for three external legs, we will prove the validity of  \eqref{eq:FF3MHVsusy-simple} for all $n$ by induction using the  BCFW recursion relation. 

A caveat is in order here: for adjacent BCFW shifts,   \eqref{eq:FF3MHVsusy-simple} 
has  a residue at $z \rightarrow \infty$. The physical interpretation of   this behaviour is interesting and will be discussed in  Section~\ref{sec:adjBCFWshift}.  On the other hand, $\F_{\T_3,n}$ has a good large-$z$ behaviour if we perform \textit{next-to-adjacent}  shifts, which we will use in the next section to prove  \eqref{eq:FF3MHVsusy-simple} for generic $n$.


 \subsection{Proof for general $n$ from recursion relations with non-adjacent shifts}
\label{nta} 
 
\begin{figure}[htb]
\centering
\includegraphics[width=0.7\textwidth]{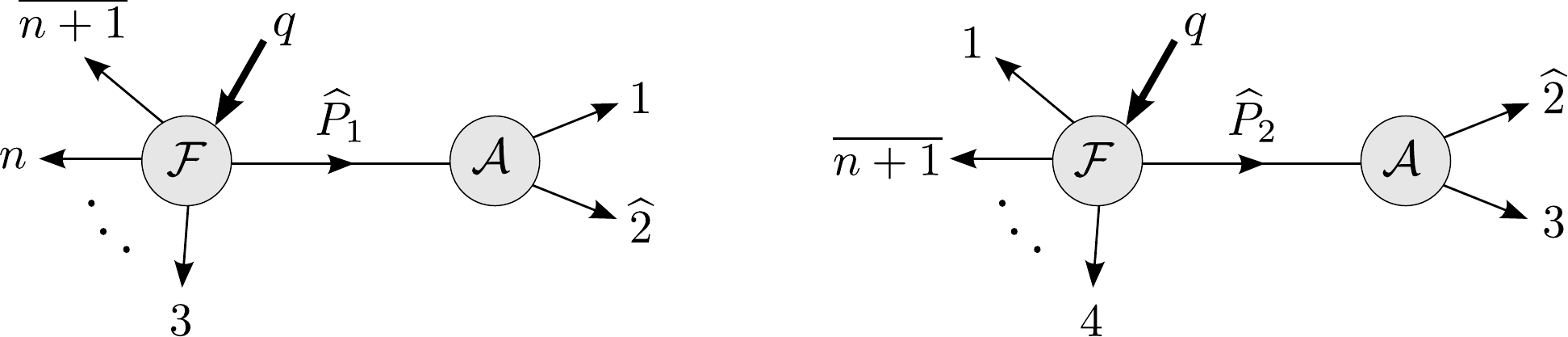}
\caption{\it The two BCFW recursive diagrams contributing to $\F^{\text{\rm MHV}}_{\T_3,n+1}$ under a next-to-adjacent  shift $(\widehat{2}, \overline{n+1})$. The amplitude on the right is $\overline{\rm MHV}$.}
\label{fig:BCFWproof}
\end{figure}

We now move on to proving  \eqref{eq:FF3MHVsusy-simple} using recursion relations. We consider the form factor with $n+1$ external particles under the following next-to-adjacent BCFW shifts, 
\begin{align}
\begin{split}
\lambda_2&\rightarrow\lambda_2+z\lambda_{n+1}\ ,
 \\
\tl_{n+1}&\rightarrow\tl_{n+1}-z\tl_{2}\ ,
 \\
\eta_{-, n+1}&\rightarrow\eta_{-, n+1}-z\eta_{-, 2}\ .
\end{split}
\end{align} 
Since in the MHV case we only have a three-particle $\MHVb$ amplitude attached to an $n$-particle MHV form factor, there are two diagrams to consider, shown in Figure \ref{fig:BCFWproof}.
\\
These are explicitly given by
\begin{align}
\F^{\text{MHV}}_{\T_3,n}( \widehat{P}_1,3,\dots,n,\overline{n+1}; q,\gamma_{+}) & \dfrac{1}{s_{12}}\A^{\MHVb}(-\widehat{P}_1,1,\widehat{2})\ , & \widehat{P}_1&=p_1+\widehat{p}_2 \, , 
\label{eq:diag1}\\
\F^{\text{MHV}}_{\T_3,n}( 1,\widehat{P}_2, 4,\dots, \overline{n,+1}; q,\gamma_{+}) &\dfrac{1}{s_{23}}\A^{\MHVb}(-\widehat{P}_2,\widehat{2},3)\ , & \widehat{P}_2&=\widehat{p}_2+p_3 \, , \label{eq:diag2}
\end{align}
where $\F^{\text{MHV}}_{\T_3,n}$ is given in \eqref{eq:FF3MHVsusy-simple} while
\beqa
\label{MHVbsupera}
\A^{\MHVb}(1,2,3)& =&\frac{\delta^{(4)}\left(\eta_{1}[23]+\eta_2[31]+\eta_{3}[12 ]\right) }{[12][23][31]}\ . 
\eeqa
It is straightforward to evaluate these two diagrams, and the corresponding results are 
\beqa
\label{eq:diag1-sol}
\text{Diag 1} & = &\F^{\text{MHV}}_{\T_2, n+1} \frac{\b{23}\b{1\,n+1}}{\b{13}\b{2\,n+1}}\left[  \sum\limits_{i>j=4}^{n+1} \b{i\,j} \eta_{-, i} \cdot \eta_{-, j}+  \sum\limits_{j=4}^{n+1}\b{3\,j}\eta_{-,3} \cdot \eta_{-, j} \right.\nonumber \\
 &+&\left. \sum\limits_{j=3}^{n+1}\b{1\,j}\left(\eta_{-, 1}+\frac{\b{2\,n+1}}{1\,n+1} \eta_{-,2}\right)\cdot \eta_{-, j} \right]\ , \cr
 \text{Diag 2} &= &\F^{\text{MHV}}_{\T_2, n+1} \frac{\b{12}\b{3\,n+1}}{\b{13}\b{2\,n+1}} \left[  \sum\limits_{i<j=4}^{n+1}\b{i\,j}\eta_{-, i} \cdot \eta_{-, j}+\b{13} \eta_{-,1} \cdot \left(\eta_{-,3}+\frac{\b{2\,n+1}}{\b{3\,n+1}}\eta_{-,1}\right) \right.
\nonumber \\
  &+ & \left.\sum\limits_{j=4}^{n+1} \b{3\,j}\left(\eta_{-,3}+\frac{\b{2\,n+1}}{\b{3\,n+1}}\eta_{-,1}\right)\cdot \eta_{-,j} \right]  \, .
\eeqa
Summing  these two contributions by collecting coefficients of $\eta_{-,i}\cdot \eta_{-,j}$, we obtain the  expected result  for the  ($n+1$)-particle form factor,
\begin{align}
\label{eq:expected}
\F^{\text{MHV}}_{\T_3,n+1}= \F^{\text{MHV}}_{\T_2, n+1} \sum\limits_{i<j=1}^{n+1} \b{i\,j} \eta_{-, i} \cdot \eta_{-, j} \ .
\end{align}
This ends the proof of our result for $\F_{\T_3, n}$ via the BCFW recursion relation. 

\subsection{A few examples of component form factors}

To conclude this section, we find it useful to present a couple of examples of component form factors. In particular we will look at the lowest component of 
$\cT_k$, which is given by the scalar operator  
\beq
\label{trfi}
\O_k(x)\, := \,\Tr \, \big[\big(\phi_{12}(x)\big)^k\big]
\ . 
\eeq
To begin with, we consider  the simple case $k=3$.  
From Feynman diagrams, it is immediate to see that at tree level the  form factor of $\O_3(x)$ is equal to one (apart from a trivial momentum conservation delta function):
\begin{align}
\begin{split}
F_{\O_3,3}(1^{\phi_{12}},2^{\phi_{12}},3^{\phi_{12}};q)\,:=&\int\!d^4x \ e^{-i q x} \ \la 1^{\phi_{12}},2^{\phi_{12}},3^{\phi_{12}} |\Tr\, \big[\big(\phi_{12}(x)\big)^3\big]|0\ra \\
=& \, \delta^{(4)}\big(q-\sum_{i=1}^3 \lambda_i\tl_i\big)\ .
\end{split}
\end{align}
From \eqref{eq:FF3MHVsusy-simple}, 
we can immediately derive the expression for the $n$-point MHV form factor with three scalars and  $n-3$ positive-helicity gluons. This is given by
\begin{equation}
\label{eq:FF3MHVboson}
F_{\O_3,n}^{\text{MHV}}(\{g^+\},a^{\phi_{12}},b^{\phi_{12}},c^{\phi_{12}};q)=\frac{\b{ab}\b{bc}\b{ca}}{\b{12}\b{23}\cdots\b{n1}}\ \delta^{(4)}\big(q-\sum_{i=1}^n \lambda_i\tl_i\big)\ ,
\end{equation}
where the  three scalars $\phi_{12}$ are at positions $a,b,c$. Notice that \eqref{eq:FF3MHVboson} scales as $(\lambda_{i})^0$ for $i \in \{a,b,c\}$ and  $(\lambda_{i})^{-2}$ for $i\notin \{a,b,c\}$ as required.
 
In fact, similar arguments can be used to write down a very concise formula for the MHV form factor of $\cO_k$ with $k$ scalars and $n-k$ positive-helicity gluons for general $k$. It contains a ratio of  Parke-Taylor factors, where in the numerator only the (ordered) scalar particle momenta appear, while the denominator is the standard Parke-Taylor expression for $n$ particles,
\begin{equation}
\label{eq:FFmMHVboson}
F_{\O_k,n}^{\text{MHV}}(\{g^+\},i_1^{\phi_{12}},i_2^{\phi_{12}},\dots, i_k^{\phi_{12}};q)=\frac{\b{i_1\,i_2}\b{i_2\,i_3}\cdots \b{i_k\,i_1}}{\b{12}\b{23}\cdots\b{n1}} \ \delta^{(4)}\big(q-\sum_{i=1}^n \lambda_i\tl_i\big)\ .
\end{equation}
The correctness of \eqref{eq:FFmMHVboson}  can easily be shown using BCFW recursion relations \cite{BCF:RecursionRelations,Britto:2005} with adjacent shifts applied to form factors \cite{Brandhuber:2010ad}.
We will not present this proof here, rather we will now consider its supersymmetric generalisation.


\section{A new recursion relation and  conjecture for the  MHV  super form factors of  $\cT_k$ }
\label{sec:adjBCFWshift}

In this section we will propose a new recursion relation for the form factors of the half-BPS supersymmetric operators $\cT_k$, shown below in \eqref{eq:recursion}. This new recursion relation is quite different from the usual BCFW recursion relation applied to form factors, in that it relates form factors of operators $\cT_k$ with different $k$. In the following we will motivate this recursion relation, whose origin lies in the presence of certain boundary terms in the usual supersymmetric BCFW recursion relation for $\cT_k$ with adjacent shifts. Following this, we will conjecture an expression for the  MHV form factors of the operators $\cT_k$  for general $k$ and show that it satisfies this new recursion relation as well as the cyclicity requirement, at some lower values of $k$ and $n$. 

\subsection{A new recursion relation for form factors}

\begin{figure}[htb]
\centering
\includegraphics[scale=0.7]{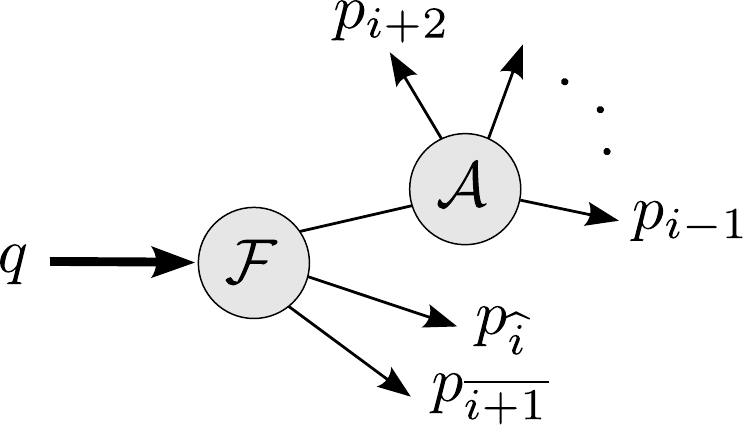}
\caption{\it Adjacent BCFW shifts on the form factor $\F_{\T_3,n}$ generate a constant contribution at $z=\infty$ when the shifted legs are directly attached to the operator, as in the diagram above.}
\label{fig:adjBCFWshift}
\end{figure}

As  observed in Section \ref{sec:susyff}, the tree-level expression \eqref{eq:FF3MHVsusy-simple}  develops a non-vanishing large-$z$ behaviour under an adjacent BCFW shift. The reason for this is that, in the case of  an adjacent shift, there will always be a Feynman diagram in which the deformed legs will be directly attached to the operator, as in Figure \ref{fig:adjBCFWshift}. In this case, the $z$-dependence will completely drop out,  leaving  a constant term at $z=\infty$. 
\\
In the case of $\T_3$, we can circumvent this problem  by using a next-to-adjacent shift, for which there is no pole at infinity. Indeed, this is the strategy we followed in Section \ref{nta} in order to determine the form factors of $\T_3$ from recursion relations. The situation is  worse   for the operators $\T_{k}$ with  $k>3$: one can convince oneself that even with non-adjacent  shifts, the diagram where the $z$-dependence drops out cannot be eliminated.

This feature impels us to look for other means to study form factors of $\T_{k}$ for general $k$. Fortunately,  the exploration of the boundary term for adjacent BCFW shifts brought to our attention an intriguing recursion relation relating the MHV form factors $\F_{\T_k,n}$, $\F_{\T_{k-1},n-1}$ and $\F_{\T_k,n-1}$, as we will now  discuss.

The claim is that  the residue  at $z \rightarrow \infty$, $R^{\rm MHV}_{\T_k,n}$,     of the $n$-particle form factor $\F^{\text{MHV}}_{\T_{k},n}$  shifted according to the  BCFW shifts 
\beqa \label{BCFWshift}
\begin{split}
\lambda_{n} &\rightarrow& \lambda_{n} -z \lambda_{n-1} \, , \\
\tl_{n-1} &\rightarrow &\tl_{n-1} + z \tl_{n} \, ,  \\   
\eta_{n-1} &\rightarrow &\eta_{n-1} + z \eta_{n} \, , 
\end{split}
\eeqa 
is given by 
\begin{align} \label{residue}
R^{\rm MHV}_{\T_k,n}  \ = \   (\eta_{-, n})^2 \widetilde{\F}^{\text{MHV}}_{\T_{k-1},n-1}(1,\dots,n-1;q,\gamma_{+}) \, .  
\end{align}
In this and in following equations, $\widetilde{\F}$ is the form factor $\F$ with the momentum and supermomentum conservation delta functions stripped off. 
For the case $k=3$,  we can confirm  this by simply using our result for  $\F^{\text{MHV}}_{\T_{3},n}$ given in \eqref{eq:FF3MHVsusy-simple}. By performing   the BCFW shift \eqref{BCFWshift}, and using supermomentum conservation, we find that the  residue at $z \rightarrow \infty $ is, on the support of the delta functions,   
\begin{eqnarray}
\begin{split}
R^{\rm MHV}_{\T_3,n}  &= 
\frac {  \sum_{i=1}^{n-2} \b{i\,n-1} \eta_{-, i}\cdot \eta_{-, n}}{\b{12}\b{23}\cdots \b{n-1\, n} \b{n-1 \ 1}} 
\ = \frac{(\eta_{-,n})^2
} {\b{12}\b{23}\cdots  \b{n-2 \, n-1} \b{n-1 \, 1}}  \, ,
\end{split}
\end{eqnarray}
which is indeed simply $(\eta_{-,n})^2 \times \widetilde{\F}^{\text{MHV}}_{\T_{2},n-1}(1,\dots, n-1;q,\gamma_{+})$. 

Conceptually, this result is very interesting since it shows that the form factors of the operator $\T_k$ are related to the form factors of the operator $\T_{k-1}$ in a simple manner. In practice, \eqref{residue}  allows us to determine the $n$-particle form factor $\F^{\text{MHV}}_{\T_{k},n}$ from the $(n-1)$-~particle form factors $\F^{\text{MHV}}_{\T_{k},n-1}$ and $\F^{\text{MHV}}_{\T_{k-1},n-1}$ in the following way: 
\begin{align}
\label{eq:recursion}
\begin{split}
\widetilde{\F}^{\text{MHV}}_{\T_{k},n}(1, \ldots, n; q,\gamma_{+}) &=   {\b{n-1 \, 1} \over \b{n-1 \, n} \b{n \, 1 } } \widetilde{\F}^{\text{MHV}}_{\T_{k},n-1}(1^\prime,2, \ldots, n-2, (n-1)^\prime; q,\gamma_{+})
 \\ \cr
&+ (\eta_{-, n})^2 \ \widetilde{\F}^{\text{MHV}}_{\T_{k-1},n-1}(1,\ldots, n-1; q,\gamma_{+}) \ ,  
\end{split}
\end{align}
where we have solved the BCFW diagram in the inverse soft form \cite{ArkaniHamed:2009si, BoucherVeronneau:2011nm, Nandan:2012rk, ArkaniHamed:2012nw}  -- indeed  the first term in \eqref{eq:recursion} simply adds particle $n$ to the $(n-1)$-particle form factor $\F^{\text{MHV}}_{\T_{k},n-1}$ with a soft factor. To maintain momentum conservation, we need to shift the legs adjacent to $n$, i.e. $(n-1)^\prime$ and $1^\prime$, with the corresponding shifted spinors given by
\begin{eqnarray}
\begin{split}
\tl_{(n-1)'} &= &\tl_{n-1} + \frac{ \b{n \, 1} } {\b{ n-1 \, 1} } \tl_{n} \,  , \qquad 
\tl_{1'} = \tl_{1} + \frac{ \b{n \, n-1} } {\b{ 1 \, n-1} } \tl_{n}\, ,  \\ 
\eta_{(n-1)'} &= &\eta_{n-1} + \frac{ \b{n \, 1} } {\b{ n-1 \, 1} } \eta_{n} \, ,  \qquad \
\eta_{1'} = \eta_{1} + \frac{ \b{n \, n-1} } {\b{ 1 \, n-1} } \eta_{n}  \, .
\end{split}
\end{eqnarray}
The second term in \eqref{eq:recursion} is again an $(n-1)$-particle form factor, but now for the operator $\T_{k-1}$. The factor $(\eta_{-, n})^2$ ensures that the fermionic degree of the expression is correct. The recursion relation may be recast into a slightly different form by removing the Parke-Taylor prefactor, 
\begin{align}
\label{recursion2}
\begin{split}
f_{\T_{k},n}(1, \ldots, n) & =   f_{\T_{k},n-1}(1',2, \ldots, n-2, (n-1)')
 \\ \cr
&+ (\eta_{-, n})^2 \ f_{\T_{k-1},n-1}(1,\ldots, n-1)  { \b{n-1 \, n} \b{n \, 1 } \over \b{n-1 \, 1}  }
\ , 
\end{split}
\end{align}
where we have defined $f_{\T_{k},n}(1, \ldots, n)$ as
\begin{equation}
\label{eq:little-f}
\F^{\text{MHV}}_{\T_{k},n}(1, \ldots, n; q,\gamma_{+}) \ := \ \F^{\text{MHV}}_{\T_{2},n}(1, \ldots, n; q,\gamma_{+})\  f_{\T_{k},n}(1, \ldots, n) \, .
\end{equation}
Given the fact that the form factors of the operator $\T_2$ are  simply given by the well-known Parke-Taylor formula, and the $k$-point form factor of the operator $\T_k$ is just one (or, in a supersymmetric language, $\prod^k_{i=1} (\eta_{-,i})^2$, the recursion relation \eqref{eq:recursion} fully determines all MHV form factors for any operator $\T_k$. Indeed, in the next section  we will propose an explicit solution to the recursion relation for the form factor $\F^{\rm MHV}_{\T_k,n}$. 


Having found a novel recursion relation \eqref{eq:recursion} for MHV super form factors, we would like to study how to generalise it to non-MHV helicities. Non-adjacent  shifts work as well for the non-MHV form factor of the operator $\T_3$,  which in principle fully determines all form factors of this operator. We can use them in order to derive the expression of non-MHV form factors, of which we can then study the large-$z$ behaviour under adjacent  shifts. 
\begin{figure}[htb]
\centering
\includegraphics[width=0.85\textwidth]{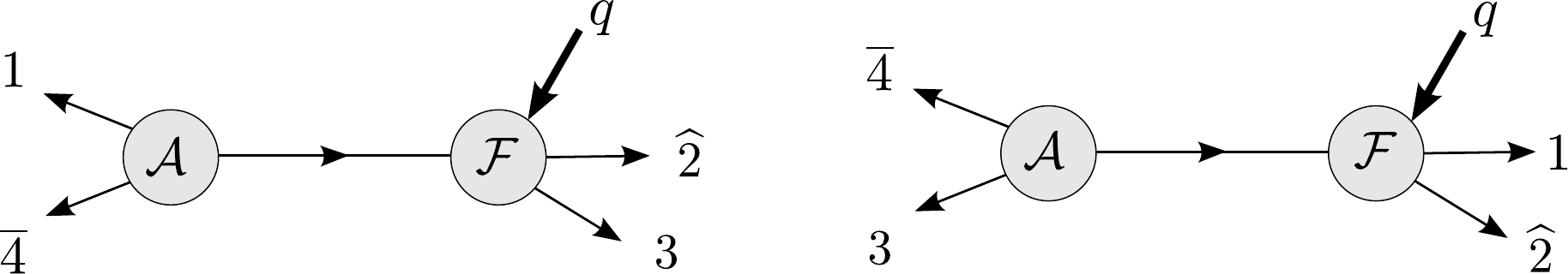}
\caption{\it Recursion relation for the  four-point NMHV form factor of $\T_3$. The amplitude on the left is MHV.}
\label{fig:4ptNMHV}
\end{figure}

The simplest non-MHV form factor is the NMHV four-particle form factor of $\T_3$. From the recursion relation with non-adjacent BCFW shifts on legs $2$ and $4$ given in Figure \ref{fig:4ptNMHV}, we find the following result, 
\begin{align}
\label{sss}
\begin{split}
\F^{\text{NMHV}}_{\T_{3},4}
&=\delta^{(4)}\big(\gamma_+ - \sum_i \lambda_i \eta_{+, i} \big) \delta^{(4)}
\big(\sum_i \lambda_i \eta_{-, i} \big)
{\delta^{(4)}\big( [23]\eta_4 + [34] \eta_2 + [42] \eta_3 \big)
(\eta_{-,1})^2 \over [23][34][42] s^2_{234} } 
 \\ 
&- (1 \leftrightarrow 3 ) \\  
&=
\delta^{(4)}(\gamma_+ - \sum_i \lambda_i \eta_{+, i} ) \prod^4_{i=1} (\eta_{-,i})^2
\Big[ 
{\delta^{(2)}( [23]\eta_{+,4} + [34] \eta_{+,2} + [42] \eta_{+,3} ) \over [23][34][42] }  \\ 
&- (1 \leftrightarrow 3 ) 
\Big] \, .
\end{split}
\end{align}
If we expand the fermionic delta function $\delta^{(2)}\big( [23]\eta_{+,4} + [34] \eta_{+,2} + [42] \eta_{+,3} \big)$, we find  non-trivial  agreement with the result \eqref{NMHV4pt1} that we will derive later using MHV rules. 

Having obtained \eqref{sss}, we can find its behaviour under adjacent BCFW shifts. Choosing the adjacent shifts 
\begin{eqnarray}
\begin{split}
\lambda_1 &\rightarrow& \lambda_1 - z\, \lambda_2 \, ,  \\
\tl_2 &\rightarrow& \tl_2  + z\, \tl_1 \, ,  \\
\eta_2 &\rightarrow &\eta_2 + z \,\eta_1 \, ,
\end{split}
\end{eqnarray}
we find that the residue of $\F^{\text{NMHV}}_{\T_{3},4}$ at large $z$ is given by
\begin{eqnarray}
\label{com}
\begin{split}
R^{\text{NMHV}}_{\T_{3},4} &= \ 
\delta^{(4)}\big(\gamma_+ - \sum\limits_{i=1}^4 \lambda_i \eta_{+,i}\big) \delta^{(4)}\big(\sum\limits_{i=1}^4 \lambda_i \eta_{-,i}\big)
{z^4\,\delta^{(4)}( [13]\eta_4 + [34] \eta_1 + [41] \eta_3 )  \over z^2 \, [13][34][41] (2\, q \cdot p_{\widehat{1}})^2 }\, (\eta_{-,1})^2\\
&=\ 
 {q^4 \over \langle 2| q |1]^2 } (\eta_{-, 1})^2 \times \widetilde{\F}^{\text{NMHV}}_{\T_{2},3}(3,4,1; q,\gamma_{+})  \, . 
\end{split}
\end{eqnarray}
In the last step we related  the residue of the NMHV form factor of the operator $\T_3$  at infinity with the 
NMHV form factor of $\T_2$, similarly to the case of the MHV form factors considered earlier.  
From \eqref{com} we see that the structure of this boundary term is more complicated than in the MHV case. 
It would be of interest to understand  this  boundary term for a general non-MHV form factor.


\subsection{Supersymmetric MHV  form factors of $\T_k$}

In this section we will propose a solution to the recursion relation \eqref{eq:recursion} for the form factor $\F^{\text{MHV}}_{\T_k,n}$.  We begin by considering the  case $k=4$. After computing a few simple examples by using the recursion relation \eqref{eq:recursion}, a clear pattern appears for $\F^{\text{MHV}}_{\T_4,n}$, which is given by 
\begin{align}
\label{eq:O4}
\F^{\text{MHV}}_{\T_4,n}\ = \ \F^{\text{MHV}}_{\T_2,n} \sum\limits_{1\leq i\leq j}^{n-3} \sum\limits_{j< k\leq l}^{n-2} (2-\delta_{ij})(2-\delta_{kl}) \frac{\b{n\,i}\b{j\,k}\b{l\,n-1}}{\b{n-1\,n}}(\eta_{-, i}\cdot \eta_{-,j})(\eta_{-,k}\cdot \eta_{-,l})\, .
\end{align}
This is clearly  a generalisation of the $k=3$ case for $\F^{\text{MHV}}_{\T_3,n}$ considered  in \eqref{eq:FF3MHVsusy}. 

Further generalisation of $\F^{\text{MHV}}_{\T_3,n}$ and $\F^{\text{MHV}}_{\T_4,n}$ leads to a proposal for $\F^{\text{MHV}}_{\T_k,n}$ for arbitrary $k$. 
In general we will have $2(k-2)$ nested sums with  fermionic degree  $2(k-2)$ in  $\eta_{-}$ (besides the delta function of supermomentum conservation).
Our conjecture for $\F^{\text{MHV}}_{\T_k,n}$ is 
\begin{align}
\label{conj}
\begin{split}
\F^{\text{MHV}}_{\T_k,n}  =\; &\F^{\text{MHV}}_{\T_2,n} 
\sum\limits_{1\leq i_{1a}\leq i_{1b}}^{n-k+1}  \sum\limits_{i_{1b}< i_{2a}\leq i_{2b}}^{n-k+2} \cdots \!\!
\sum\limits_{i_{(k-3)b} < i_{(k-2)a} \leq i_{(k-2)b}}^{n-2} 
\\ \times &\; C_{i_{1a},i_{1b},i_{2a},i_{2b}, \cdots,i_{(k-2)a},i_{(k-2)b}}
\prod^{k-2}_{\alpha=1 } (\eta_{-, i_{\alpha a}}\cdot\eta_{-, i_{\alpha b}})
\ ,
\end{split}
\end{align}
where the coefficient  $C_{i_{1a},i_{1b},i_{2a},i_{2b}, \cdots,i_{(k-2)a},i_{(k-2)b}}$ is a natural generalisation of the coefficient in \eqref{eq:O4},
\beqa
\begin{split}
 &C_{i_{1a},i_{1b},i_{2a},i_{2b}, \cdots,i_{(k-2)a},i_{(k-2)b}} = \cr
& 
\left( \prod^{k-2}_{\alpha=1 } (2-\delta_{i_{\alpha a}i_{ \alpha b}}) \right)
 \frac {\b{n\,i_{1a}}\b{i_{1b} \,i_{2a}}\cdots \b{i_{(k-3)b}\,i_{(k-2)a}} \b{i_{(k-2)b}\,n-1}}
{\b{n-1 \,n}}  \, .
\end{split}
\eeqa
In the summations in \eqref{conj} we sum over pairs of indices $i_{\alpha  a}$, $i_{\alpha  b}$, for $\alpha =1, \ldots , k-2$. 
We have  compared   \eqref{conj} to the result obtained from  the recursion relation \eqref{eq:recursion} and agreement has been found for all cases we have checked, namely $k\leq 6,\,n\leq 7$.   

We would like to stress that,  unlike the case of the recursion for the form factor $\F_{\T_3,n}$ with  non-adjacent shifts, 
the recursion relation \eqref{eq:recursion} is a conjecture, hence it is important to check the correctness of the resulting $\F_{\T_k,n}$ in \eqref{conj}, obtained from studying  \eqref{eq:recursion}. 
One non-trivial test consists in checking  the cyclicity of the result. 
In Appendix \ref{app:cyclicity} we prove that our result  for $\F_{\T_4,n}^{\rm MHV}$ indeed enjoys this symmetry in a very non-trivial way. Unfortunately we have not been able to prove the cyclicity of $\F_{\T_k,n}$ for arbitrary $k$, however we have checked various cases for $k\leq 6$ with Mathematica and found that the required symmetry is indeed present. 
The proof of $\F_{\T_4,n}$ and these checks  lend support to our conjectured recursion relation \eqref{eq:recursion} and  solution \eqref{conj}.


\section{MHV rules for $\F_{\T_k,n}$}
\label{sec:MHVrules}

In \cite{Brandhuber:2011tv}, MHV rules for the form factor of the stress-tensor multiplet operator were constructed. Here we show in a number of concrete applications that these MHV rules can directly be extended to the form factors of the operators $\cT_k$ with $k>2$. In this approach, the usual MHV vertices of \cite{Cachazo:2004kj} are augmented by a new set of vertices obtained by continuing off-shell the holomorphic form factor expression for  $\cF_{\cT_k,n}^{\rm MHV}$ 
using the same prescription as in \cite{Cachazo:2004kj}. In the following we will illustrate the application of this technique by computing a few examples, but we comment that the approach can be used in general to obtain form factors with higher MHV degree and, following \cite{Brandhuber:2004yw}, number of loops. 

\subsection{Four-particle bosonic NMHV form factor}

\begin{figure}[htb]
\centering
\includegraphics[width=0.8\textwidth]{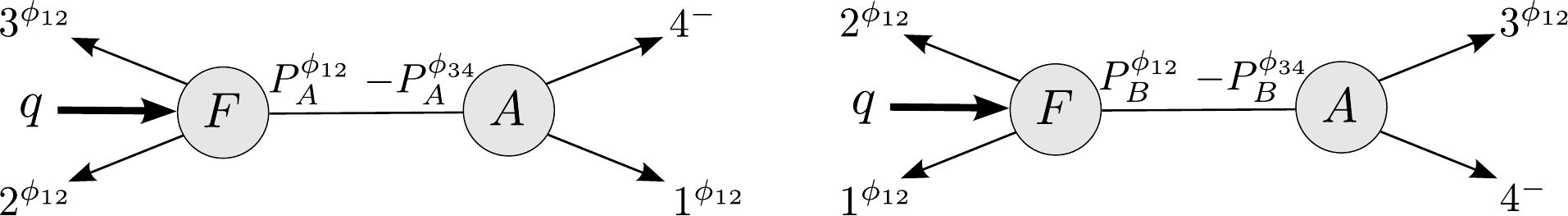}
\caption{\it Expansion of $ F^{\text{\rm NMHV}}(1^{\phi_{12}},2^{\phi_{12}},3^{\phi_{12}},4^-; q) $ using MHV rules.}
\label{fig:NMHV}
\end{figure}

As a first example, we  consider the bosonic form factor $ F^{\text{NMHV}}(1^{\phi_{12}},2^{\phi_{12}},3^{\phi_{12}},4^-; q) $ and compute it with MHV rules.  There are two diagrams that contribute to this, shown in Figure \ref{fig:NMHV}.
\\
These result in the respective expressions
\begin{align}
\label{eq:MHV41}
\big[ F^{\text{NMHV}}(1^{\phi_{12}},2^{\phi_{12}},3^{\phi_{12}},4^-;q)\big]^{(1)}  \ &= \ F^{\text{MHV}}(2^{\phi_{12}},3^{\phi_{12}},P_A^{\phi_{12}};q)\,A^{\text{MHV}}(-P_A^{\phi_{34}},4^-,1^{\phi_{12}}) \\
 \label{eq:MHV42}
 \big[ F^{\text{NMHV}}(1^{\phi_{12}},2^{\phi_{12}},3^{\phi_{12}},4^-;q)\big]^{(2)} \ &= \ 
 F^{\text{MHV}}(1^{\phi_{12}},2^{\phi_{12}},P_B^{\phi_{12}};q)\,A^{\text{MHV}}(-P_B^{\phi_{34}},3^{\phi_{12}},4^-)\ , 
\end{align}
with
\begin{align}
P_A\ &= \ p_1+p_4\ , \qquad  |A\ra \,=\,(p_1+p_4)|\xi]\ ,\nonumber \\
P_B\ &=\ p_3+p_4 \ , \qquad  |B\ra \, =\,(p_3+p_4)|\xi]\ , 
\end{align}
where $|\xi]$ is the reference spinor used in the  off-shell continuation needed in order to define spinors associated to the internal momenta $P_{A,B}$\cite{Cachazo:2004kj}. 
A crucial check of the correctness of the procedure is to confirm that the final answer for an amplitude or form factor evaluated with MHV diagrams is independent of the choice of the reference spinor $|\xi]$.

Using the fact that  that $F(a^{\phi_{12}},b^{\phi_{12}},c^{\phi_{12}};q) = 1 $ (omitting a delta function of momentum conservation), the first contribution \eqref{eq:MHV41} is simply given by 
\begin{align}
\frac{1}{\b{14}[14]}\times \frac{\b{A4}^2\b{14}^2}{\b{A1}\b{A4}\b{41}}=-\frac{\b{4A}}{[14]\b{A1}}= -\frac{\la 4|1|\xi]}{[14][\xi|4|1\ra}\ .
\end{align}
Analogously, the second contribution \eqref{eq:MHV42} is 
\begin{align}
\frac{1}{\b{34}[34]}\times \frac{\b{4B}^2\b{34}^2}{\b{34}\b{4B}\b{B3}}=\frac{\b{4B}}{[34]\b{B3}}= \frac{\la 4|3|\xi]}{[34][\xi|4|3\ra}\ .
\end{align}
Summing these, we get
\begin{align}
\frac{\la 4|3|\xi] [\xi|4|1\ra  [14]  - \la 4|1|\xi][34][\xi|4|3\ra }{[14][\xi|4|1\ra [34][\xi|4|3\ra }= \frac{[\xi| 
{p}_4 \, {p}_1 \, {p}_4 \, {p}_3 |\xi]-[\xi| {p}_4 \,  {p}_3 \, {p}_4 \, {p}_1 |\xi]}{[\xi 4] \b{43} [34] [41] \b{14}[4 \xi]}\ .
\end{align}
The numerator can be rewritten as
\beq
 [\xi 4]\b{41}\b{43}([14][3\xi]-[34][1\xi])= [\xi 4]^2\b{41}\b{43}[31]\ , 
\eeq
thus the final result is independent of the choice of $|\xi]$ and is given by
\begin{align}
F^{\text{NMHV}}(1^{\phi_{12}},2^{\phi_{12}},3^{\phi_{12}},4^-; q) = \frac{[31]}{[34][41]}\ , 
\end{align}
which is the $k$-increasing inverse soft factor, as expected.


\subsection{Four-particle super form factors}
\begin{figure}[htb]
\centering
\includegraphics[width=0.5\textwidth]{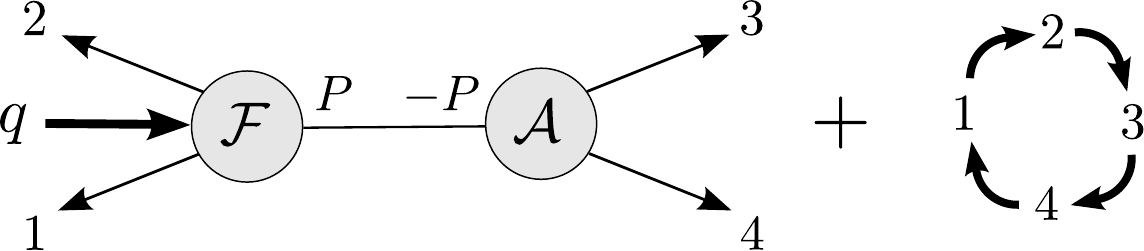}
\caption{\it Expansion of $\F^{\text{\rm NMHV}}_{\T_3,4}$ using supersymmetric MHV rules.}
\label{fig:NMHVsusy}
\end{figure}
In this section we compute the supersymmetric form factor $\F^{\text{NMHV}}_{\T_3,4}$ using MHV diagrams. The diagrams contributing are  shown in Figure \ref{fig:NMHVsusy}, which can be written as
\beq
\label{eq:4ptMHVsusy}
\F^{\text{NMHV}}_{\T_3,4} \ = \ 
\F^{\text{MHV}}_{\T_3,3}(1,2,P;q, \gamma_{+}) \dfrac{1}{\b{34}[34]}\A^{\text{MHV}}(-P,3,4)\ + \ {\rm cyclic} (1,2,3,4)\,  ,
\eeq
where 
\beq
P=p_3+p_4\ , \qquad |P\ra = (p_3+p_4)|\xi]\ , 
\eeq
while the MHV superamplitude is
\beq
\label{MHVsupera}
\A^{\rm MHV}(1, \ldots, n ) \ = \ \frac{\delta^{(8)}\big(\sum_{i=1}^{n}\lambda_i\eta_i \big)  }{\b{12}\cdots\b{n1}}\ .
\eeq 
Writing the form factor as $\dfrac{(\eta_{-, P})^2}{\b{12}^2}$, the integration over $\eta_{-, P}$ becomes simply
\beqa
\begin{split}
& \int d^2\eta_{-, P} (\eta_{-, P})^2 \delta^{(4)} \big(  \lambda_1\eta_{-, 1} +\lambda_2\eta_{-, 2} + \lambda_P\eta_{-, P}\big) \delta^{(4)} 
\big(\lambda_3\eta_{-, 3} +\lambda_4\eta_{-, 4} - \lambda_P\eta_{-, P}\big)  \\  
 = & \b{12}^2\b{34}^2\prod_{i=1}^4(\eta_{-, i})^2\ .
\end{split}
\eeqa
Integrating over $\eta_{+, P} $ gives 
\begin{align}
\begin{split}
&\int d^2\eta_{+, P} \, \delta^{(4)} \big(\gamma_+ - \lambda_1\eta_{+, 1} -\lambda_2\eta_{+, 2} - \lambda_P\eta_{+, P}\big) \delta^{(4)}\big(\lambda_3\eta_{+, 3} +\lambda_4\eta_{+, 4} - \lambda_P\eta_{+, P}\big)\\
= \ \ &\delta^{(4)}\big(\gamma_+ - \sum\limits_{i=1}^4\lambda_i\eta_{+, i} ) \delta^{(2)} \big( \b{3P}\eta_{+, 3} + \b{4P} \eta_{+, 4} \big) \ . 
\end{split}
\end{align}
substituting this into \eqref{eq:4ptMHVsusy}, we get
\begin{align}
\label{eq:NMHVsusy}
\F^{\text{NMHV}}_{\T_3,4}\  = \ \prod_{i=1}^4(\eta_{-, i})^2 \delta^{(4)}\big(\gamma_+ - \sum\limits_{i=1}^4\lambda_i\eta_{+, i} \big)  \delta^{(2)}\big( \la 3|4|\xi]\eta_{+, 3} + \la 4 |3|\xi] \eta_{+, 4} \big) \frac{1}{[\xi|4|3\ra [34]\la 4|3 |\xi]} \ .
\end{align}
We note that \eqref{eq:NMHVsusy} does not scale with the reference spinor $|\xi]$. Also, we see that all the dependence on $|\xi]$ cancels out for all coefficients of  $\eta_{+, i}\cdot \eta_{+, j}$  as follows. For the cross terms $i\neq j$, the only contribution comes from the diagram with particles $i$ and $j$ on the amplitude side, for example the diagram in Figure \ref{fig:NMHVsusy} is the only one which carries  $\eta_{+, 3}\cdot \eta_{+, 4}$  with a coefficient
\begin{equation}
\frac{\la 3|4|\xi] \la 4 |3|\xi]}{[\xi|4|3\ra [34]\la 4|3 |\xi]} = \frac{1}{[34]}\ .
\end{equation}
For the terms with $(\eta_{+, i})^2$, the contribution comes from two diagrams with particle $i$ on the amplitude side. Taking as an example the $(\eta_{+, 4})^2$ coefficient, we must also take into account the  following particular diagram, 
\beqa
\begin{split}
& \F^{\text{MHV}}_{\T_3,3}(2,3,P;q, \gamma_{+}) \dfrac{1}{\b{41}[41]}\A^{\text{MHV}}(-P,4,1)
 \\
= &   \prod_{i=1}^4(\eta_{i,-})^2 \delta^{(4)}(\gamma_{+} - \sum\limits_{i=1}^4\lambda_i\eta_{+,i} ) \delta^{(2)}\left( \la 4|1|\xi]\eta_{+,4} + \la 1 |4|\xi] \eta_{+,1} \right) \frac{1}{[\xi|1|4\ra [41]\la 1|4 |\xi]}
\end{split}
\eeqa
where 
\beq
P=p_4+p_1\ , \qquad |P\ra = (p_4+p_1)|\xi]
\ . 
\eeq
Thus, summing the coefficients of $(\eta_{+,4})^2$  we get:
\begin{align}
\frac{[\xi|1|4\ra }{[41] \la 1|4 |\xi]}+ \frac{\la 4|3 |\xi]}{[\xi|4|3\ra [34]} = \frac{[13]}{[14][43]}\ .
\end{align}
This cancellation of the reference spinor clearly happens for all $i=1,\ldots,4$. Our final result for this form factor is
\begin{align}
\label{NMHV4pt1}
\F^{\text{NMHV}}_{\T_3,4}&= \Delta^{4|4+} \prod_{i=1}^4(\eta_{-, i})^2
\times \sum_{i=1}^4 \left((\eta_{+, i})^2\frac{[i+1\,i-1]}{[i+1\,i][i\,i-1]}+\frac{\eta_{+, i} \cdot \eta_{+, i+1}}{[i\,i+1]}\right)\ ,
\end{align}
where we have defined
$
\Delta^{4|4+} :=   \delta^{(4)}\big (q - \sum\limits_{i=1}^4\lambda_i\tl_i \big)  \delta^{(4)}\big(\gamma_+ - \sum\limits_{i=1}^4\lambda_i\eta_{+, i} \big)
$.
As mentioned earlier, this result agrees with what we have obtained from non-adjacent BCFW shifts. 


\section{Super form factors of $\T_k$ at one loop}
\label{sec:1loopTk}

In this section we extend our previous analysis and compute supersymmetric form factors of $\cT_k$ at one loop. On general grounds, 
we can  expand $\F^{(1)}_{\T_k,n}$ as%
\footnote{The precise definitions of the various  triangle  and box  integrals  can be found   in Appendix~\ref{app:integrals}.}
\begin{align}
\label{eq:all-1loop}
\F_{\T_k,n}^{(1)}\ =\  -\F_{\T_k,n}^{(0)} \sum_{i=1}^n s_{i\,i+1} I_{3;i}^{1m}(s_{i\,i+1}) \, + \, \text{finite boxes} \, + \, \text{three-mass triangles}\, , 
\end{align}
where $I_{3;i}^{1m}$ is a one-mass triangle, and $s_{i\, i+1} := (p_i +p_{i+1})^2$.
We can motivate \eqref{eq:all-1loop}  by knowing that the answer should be expressed in terms of triangles and boxes (bubbles are absent since the theory is finite in the ultraviolet). Furthermore, the infrared-divergent part of  any one-loop form factor must be proportional to its  tree-level counterpart in order to guarantee the correct exponentiation of these divergences, as we will explicitly show in the next section. This explains the first term in \eqref{eq:all-1loop}.  In practice, all the infrared divergences contained in the box functions which do not contain two-particle invariants $s_{i\, i+1}$ have to cancel with corresponding divergences from one-mass triangles, leaving behind only finite boxes and a collection of one-mass triangles where the massless legs are $p_i$ and $p_{i+1}$. 

This discussion leaves room for three-mass triangles, and does not put any constraints on what finite boxes will appear. However, the form factors with MHV helicity configuration which we will consider are special in two ways:

{\bf 1.} 
Three-mass triangles are in fact absent.  This can easily be understood by counting the fermionic degree of the cut diagram. Consider a triple cut contributing to this form factor, with two amplitudes and one form factor participating to the cut. The MHV form factor $\F_{\T_k,n}$ has fermionic degree  $2( k-2) +8$, and hence one of the two superamplitudes must be a three-point $\overline{\rm MHV}$ superamplitude, so that the overall fermionic degree is $2(k-2)+8 + 8 + 4 - 4\times 3 = 2(k-2)+8$. Thus, at most two-mass triangles can be present.

{\bf 2.}  Only two-mass easy boxes can appear (or one-mass for up to three external legs), similarly to the one-loop MHV superamplitudes. The reason is the same as for the MHV superamplitudes: in order to obtain the correct fermionic degree there must be two three-point $\overline{\rm MHV}$ superamplitudes participating in the cut (the overall fermionic degree being $2(k-2)+8 + 4 + 4 + 8 - 4\times 4 = 2(k-2)+8$), and these two three-point $\overline{\rm MHV}$ superamplitudes must not be adjacent in order not to constrain the external kinematics. 
Of course already at the NMHV case we expect to find  two-mass hard, three-mass and four-mass boxes as well as three-mass triangles, as indicated in \eqref{eq:all-1loop}. 

The strategy we will follow will then consist in computing the coefficient of the {\it finite} box functions using quadruple cuts. The complete result for the one-loop MHV super form factor will then be given by the sum of these finite box functions with the one-mass triangles accounting for the expected infrared divergences. 

In the remaining part of this section we will first derive the infrared-divergent part of general one-loop form factors. Next, we will consider the simplest case, that of the Sudakov form factor, which we will compute using two-particle cuts. Finally, we will derive the expression of MHV form factors for general $n$ and $k$ using quadruple cuts.

\subsection{General  infrared-divergent structure  of form factors}

As noted in \cite{Bena:2004xu},
the infrared divergences of generic one-loop amplitudes in $\mathcal{N}=4$ SYM are  captured by a particular two-particle cut diagram where on one side of the cut there is a four-point amplitude.%
\footnote{See also \cite{Brandhuber:2009kh} for an application of the same ideas to dual conformal anomalies at one loop.} 
\begin{figure}[h]
\centering
\includegraphics[width=0.7\textwidth]{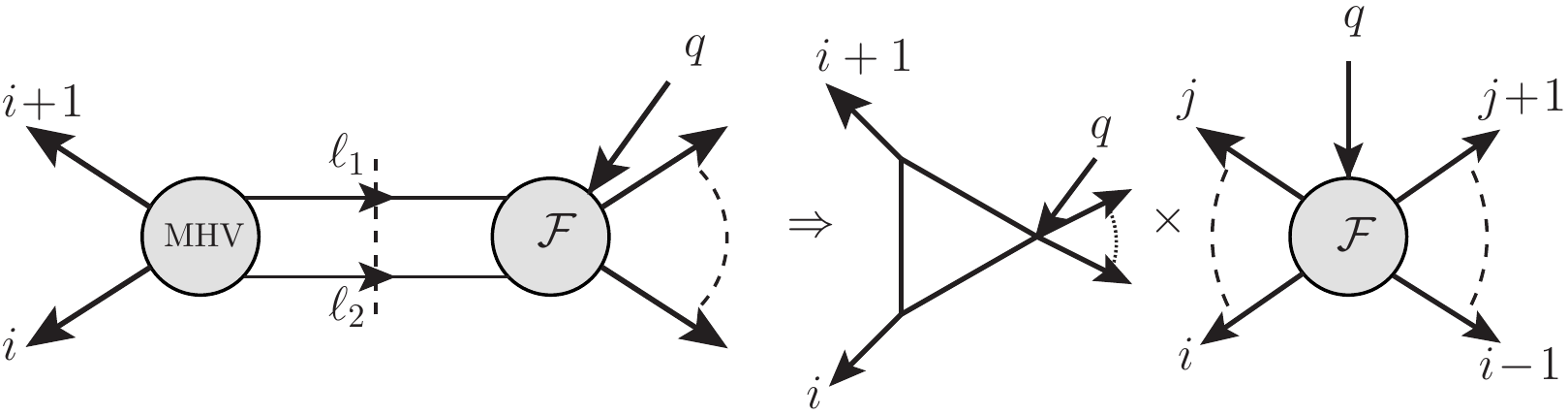}
\caption{\it The two-particle cut diagram which captures the infrared divergences of general one-loop form factor. The integration region responsible for the infrared divergences is the forward-scattering region, where   $\ell_1 \rightarrow -p_{i+1}$ and $\ell_2 \rightarrow -p_i$.}
\label{fig:IRcut}
\end{figure}
The same is true for form factors, and their infrared divergences are fully captured by a two-particle cut diagram where the participating amplitude is a four-point amplitude. Infrared divergences arise from a particular region in the space of internal momenta  $\ell_1$ and $\ell_2$, namely the forward scattering region (see Figure \ref{fig:IRcut}). Indeed, when  $\ell_1 \rightarrow -p_{i+1}$, the four-point kinematics also forces  $\ell_2 \rightarrow -p_{i}$, and this creates a simple pole which is responsible for the infrared divergence of the amplitudes.  Following the same proof as in \cite{Brandhuber:2009kh}, it is very easy to show that in the limit $\ell_1 \rightarrow -p_{i+1}$ and $\ell_2 \rightarrow -p_{i}$, the two-particle cut in question can be uplifted to a one-mass triangle integral multiplied by the tree-level form factor. Summing over all the channels, we obtain the leading infrared divergence of generic form factors%
\footnote{In writing the second equality we have dropped a factor of $e^{ \gamma_{\rm E} \epsilon}\, r_\Gamma = 1+ \cO(\epsilon^2)$, where $r_\Gamma$ is defined in~\eqref{rgamma}.} 
\begin{align}
\label{eq:all-1loopIR}
\F_{\T_k,n}^{\rm IR} \ = \ -\F_{\T_k,n}^{(0)} \sum_{i=1}^n s_{i\,i+1} I_{3;i}^{1m}(s_{i\,i+1})  \ = \   \F_{\T_k,n}^{(0)}\
\sum_{i=1}^n { (-s_{i\,i+1} )^{-\epsilon} \over \epsilon^2 }   \, .
\end{align}
We also note that in $\cN=4$ SYM  the  above result is correct at leading and subleading order in   $1/\epsilon$. %
\subsection{Three-point super form factor of $\T_3$}

\begin{figure}[htb]
\centering
\includegraphics[scale=0.7]{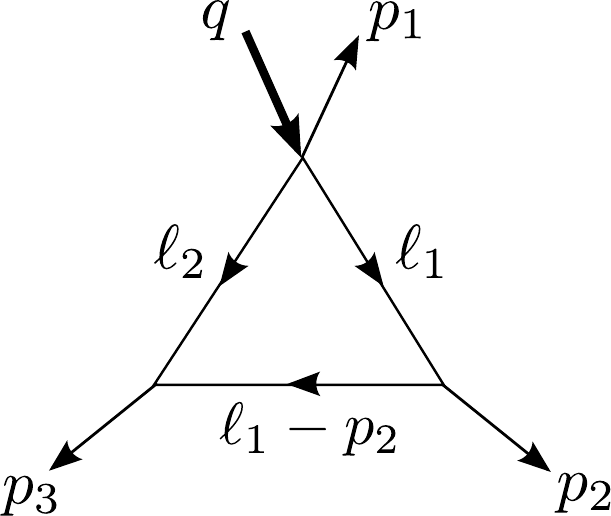}
\caption{\it The result for the $(q-p_1)^2$ cut of the one-loop Sudakov form factor of $\T_3$.}
\label{fig:1loop3}
\end{figure}

As a warm-up, we start by computing the simplest form factor at one loop, namely the Sudakov form factor.
\begin{figure}[h]
\centering
\includegraphics[width=0.4\textwidth]{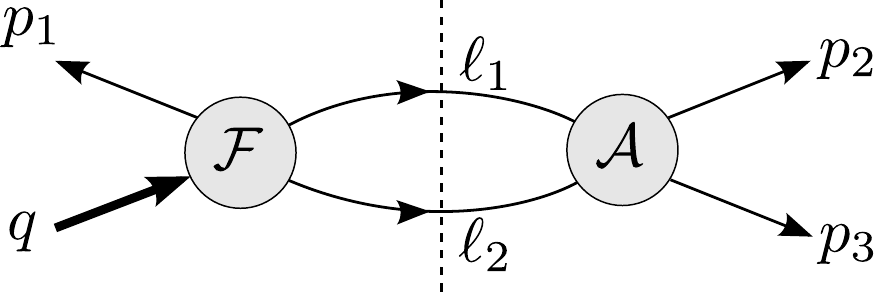}
\caption{\it $(q-p_1)^2$ two-particle cut for the Sudakov super form factor of $\T_3$.}
\label{fig:cut3}
\end{figure}\\
The cut of $\F^{(1)}_{\T_3,3}$ across the $(q-p_1)^2$ channel, shown in Figure \ref{fig:cut3}, is given by
\beqa
\label{eq:1loop3}
\begin{split}
 &  \int\!\! d\text{LIPS}(\ell_1,\ell_2;P)\, \F_{\T_3,3}(1,\ell_1,\ell_2;q,\gamma_{+})\ \A^{\text{MHV}}(-\ell_1,2,3,-\ell_2) \\  
 =\, &  \int\!\! d\text{LIPS}(\ell_1,\ell_2;P) \frac{(\eta_{-,1})^2}{\b{\ell_1\,\ell_2}^2}\frac{\b{\ell_1\,\ell_2}^4}{\b{23}\b{3\,\ell_2}\b{\ell_2\,\ell_1}\b{\ell_1\,2}}  \\
 =\, &\frac{(\eta_{-,1})^2}{\b{23}^2}  \int\!\! d\text{LIPS}(\ell_1,\ell_2;P) \frac{\b{23} [3\,\ell_2] \b{\ell_2\,\ell_1} [\ell_1\,2]}{4(p_3\cdot \ell_2)(p_2\cdot \ell_1)} \ , 
\end{split}
\eeqa
where $P=q-p_1$, the MHV superamplitude is given in \eqref{MHVsupera}, and $d\text{LIPS}(\ell_1,\ell_2;P) $ stands for the usual two-particle phase space measure.
 Using $\ell_1 + \ell_2 = p_2+p_3$, the numerator of \eqref{eq:1loop3} can be written as  
$2 s_{23}(p_2\cdot \ell_1)  $, 
 thus the result is a one-mass triangle with massive corner $P$, as shown in Figure \ref{fig:1loop3}.
There is no ambiguity in lifting this cut to a full integral \cite{Kosower:1999xi}. Summing over the contribution of all cuts we arrive at  
the complete result for $\F^{(1)}_{\T_3,3}$, 
\begin{align}
\label{eq:1loop3ff}
\F^{(1)}_{\T_3,3}\ = \ 
\F^{(0)}_{\T_3,3} \, \sum_{i=1}^3 { ( -  s_{i\,i+1} )^{-\epsilon} \over \epsilon^2} \ . 
\end{align}
We mention that this  one-loop Sudakov form factor of $\Tr\big[ (\phi_{12})^3\big] $ was computed earlier in \cite{Bork:2010wf} using supergraphs,  and our result agrees with theirs.

\subsection{$n$-point MHV super form factors of $\T_3$}

As stated earlier, we only need to compute the quadruple cut diagrams of the one-loop MHV super form factor of $\T_3$. The final result will then be expressed as a sum of the infrared-divergent expression \eqref{eq:all-1loopIR} plus finite two-mass easy boxes, whose coefficients we are going to determine now using supersymmetric quadruple cuts  \cite{Drummond:2008bq}. 

\begin{figure}[htb]
\centering
\includegraphics[width=0.28\textwidth]{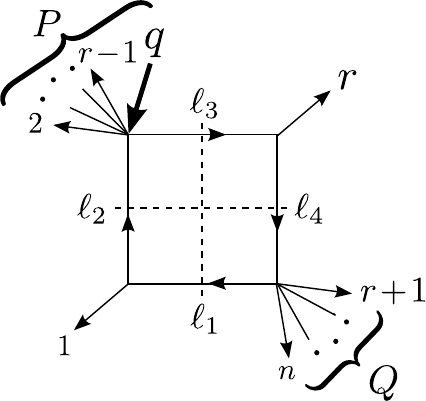}
\caption{\it Quadruple cut of the super form factor $\F_{\T_3,n}^{\text{\rm MHV}(1)}$.}
\label{fig:quadcut}
\end{figure}

The two-mass easy quadruple cuts we consider are shown in Figure \ref{fig:quadcut}, where for convenience we label the massless legs $1$ and $r$.
\\
The coefficient of the corresponding box is given by 
\begin{align}
\begin{split}
\label{eq:qc-coeff}
\cC (1, P, r, Q) \ = \  {1\over 2} \sum_{\mathcal{S}_{\pm}} \int\!\prod_{i=1}^4 d^4 \eta_i \ 
 \F^{\text{MHV}}_{\T_3,r}(2,\dots,r-1,\ell_3,-\ell_2;q,\gamma_{+})\times \A^{\MHVb}(-\ell_3,r,\ell_4)& \\
\times \, \A^{\text{MHV}}(-\ell_4,r+1,\dots,n,\ell_1)\times \A^{\MHVb}(-\ell_1,1,\ell_2)& \ , 
\end{split}
\end{align}
where the sum is over the solutions to the cut equations. Since only one solution to the cut contributes, one can drop the sum over $\mathcal{S}_{\pm}$ in \eqref{eq:qc-coeff}, and a factor of 1/2 is left over.
The form factor $ \F^{\text{MHV}}_{\T_3,r}$  is given in \eqref{eq:FF3MHVsusy}, and the $\text{MHV}$ and $\overline{\rm MHV}$ superamplitudes entering this expression are 
given in \eqref{MHVsupera} and  \eqref{MHVbsupera}, respectively. 
Because of the presence of $\MHVb$ three-particle amplitudes on the massless corners, we have
\begin{equation}
\label{eq:proplambdas}
\lambda_{\ell_3}\propto\lambda_{\ell_4}\propto\lambda_r\ ,\qquad\qquad \lambda_{\ell_1}\propto\lambda_{\ell_2}\propto\lambda_1\ .
\end{equation}
Using  the delta functions contained in the $\MHVb$  and MHV amplitudes, together with the conditions \eqref{eq:proplambdas} one can quickly determine the fermionic variables associated to the internal supermomenta, 
\begin{equation}
\eta_{\ell_4}=  \sum_{i=r+1}^n \frac{\b{1\,i}}{\b{1\,\ell_4}}\,\eta_i \ , \quad
\eta_{\ell_1}= - \sum_{i=r+1}^n \frac{\b{i\,r}}{\b{\ell_1\,r}}\,\eta_i\ , 
\end{equation}
and 
\begin{align}
\label{eq:etal3b}
\eta_{\ell_3} = \frac{[\ell_4\,\ell_3]}{[\ell_4\,r]}\,\eta_r +\sum_{i=r+1}^n \frac{\b{1\,i}[\ell_3\,r]}{\b{1\,\ell_4}[\ell_4\,r]}\,\eta_i \ , \\
\label{eq:etal2b}
\eta_{\ell_2}= \frac{[\ell_2\,\ell_1]}{[1\,\ell_1]}\,\eta_r +\sum_{i=r+1}^n \frac{\b{1\,r}[1\,\ell_2]}{\b{\ell_1\,r}[1\,\ell_1]}\,\eta_i\ . 
\end{align}
Integrating out the internal $\eta$ variables produced  the two expected supermomentum conservation delta functions 
$\delta^{4} ( \gamma_+ - \sum_{i=1}^n \lambda_i \eta_{+,i}) \, \delta^{4}  (  \sum_{i=1}^n \lambda_i \eta_{-,i}) $ as well as a Jacobian 
\begin{equation} 
\label{eq:jac1}
J\ =\ (\b{\ell_1\,\ell_4}[r\,\ell_4][\ell_1\,1])^4\, =\,   [1|\ell_1\,\ell_4|r]^4\ .
\end{equation}
Let us now manipulate the Parke-Taylor prefactors coming from \eqref{eq:qc-coeff} together with \eqref{eq:jac1}:
\begin{align}
\begin{split}
&\frac{1}{\b{\ell_2\,2}\ldots\b{r-1\,\ell_3}\b{\ell_3\,\ell_2}} \times \frac{1}{[\ell_3\,r][r\,\ell_4][\ell_4\,\ell_3]} \\ 
\times \ & \frac{1}{\b{\ell_4\,r+1}\dots\b{n\,\ell_1}\b{\ell_1\,\ell_4}} \times \frac{1}{[\ell_1\,1][1\,\ell_2][\ell_2\,\ell_1]} \times (\b{\ell_1\,\ell_4}[r\,\ell_4][\ell_1\,1])^4
\\
\label{eq:ptman1}
= \ & \text{PT}_n [1|\ell_1\,\ell_4|r]^3 \frac{\b{n1}\b{12}\b{r-1\,r}\b{r\,r+1}}{\la r-1|\ell_3\,\ell_4|r+1\ra \la 2|\ell_2\,\ell_1|n\ra [ 1|\ell_2\,\ell_3|r]}\ , 
\end{split} 
\end{align}
where $\text{PT}_n:=   1 / ( {\b{12}\b{23}\cdots\b{n1}})$.
This expression can be considerably simplified by using momentum conservation and  the replacements  \eqref{eq:proplambdas} inside expressions which are homogeneous functions of degree zero of the spinors associated to the cut loop momenta. In this way one can rewrite this product of amplitudes as 
\begin{align}
-\text{PT}_n \frac{[1\,\ell_2]\b{\ell_2\,r} [\ell_3\,r]\b{\ell_3\,1} \b{r\,r+1}}{\b{r\,r+1}}=\text{PT}_n\,  \Tr_+(\ell_2  \, p_r \,   \ell_3 \,  p_1) \ .
\end{align}
Using again momentum conservation and $(p_1\cdot\ell_2)=0$ we can rewrite the trace as 
\begin{align}
\begin{split}
&\Tr_+(\ell_2  \, p_r \,   \ell_3 \,  p_1) \ = \ \Tr_+(Q\, p_r \, P \,  p_1) \ = \ 2 (p_1\cdot P) (p_r\cdot Q) + 2 (p_r\cdot P) (p_1\cdot Q) -s_{1r} (Q\cdot P) 
\ . 
\end{split} \nonumber \\[5pt]
\end{align}
Introducing the kinematic variables 
\begin{align}
s\ :=\ (p_r+Q)^2\ , \qquad t\ :=\ (p_r+P)^2\ , 
\end{align}
we can write  $s_{1r}\ =\  - (s+t-P^2-Q^2)$. With that we can finally rewrite the trace as
\begin{align}
\Tr_+(\ell_2  \, p_r \,   \ell_3 \,  p_1)  \ = \ P^2 Q^2-st\ .
\end{align}
Substituting this back into \eqref{eq:qc-coeff},  we arrive at the result for the supercoefficient, 
\beq
\label{eq:result-qc}
\cC(1,P, r, Q) \ = \ 
\,
\F_{\T_2,n}^{\text{MHV}(0)} 
 \left( P^2 Q^2-st \right)  
\  \delta\Big(\dfrac{1}{2}\, \sum\limits_{i<j=2}^{r-1}(2-\delta_{ij})\dfrac{\b{1\,i}\b{j\,r}}{\b{r\,1}}\eta_{-,i} \cdot \eta_{-,j}\Big)\ .
\eeq
We note that the delta function appearing above corresponds precisely to that of the form factor entering the quadruple cut, where we conveniently singled out the two internal loop legs (the corresponding 
spinor variables being in turn proportional to the two external momenta entering the adjacent massless corners $1$ and $r$, as per  \eqref{eq:proplambdas}). We can therefore rewrite \eqref{eq:result-qc}~as
\begin{align}
\label{ccpp}
\cC(1,P, r, Q) = \F_{\T_2,n}^{\text{MHV}(0)}   \left( P^2 Q^2-st \right) f_{\T_3,r}(2,\dots , r-1,r,1)\, , 
\end{align}
where $f_{\T_3,r}$ is defined in \eqref{eq:little-f}%
\footnote{We stress that,  in \eqref{ccpp},  we should use the form of the quantity  $f_{\T_3,r}$ (defined in  \eqref{eq:little-f}) given in \eqref{eq:FF3MHVsusy} and not \eqref{eq:FF3MHVsusy-simple}. The reason is that these two expressions are  only equivalent on the support of the delta function $\delta\big(\sum_{i=1}^rp_i-q\big)$, which is not true in this case.}.

\begin{figure}[htb]
\centering
\includegraphics[width=0.65\textwidth]{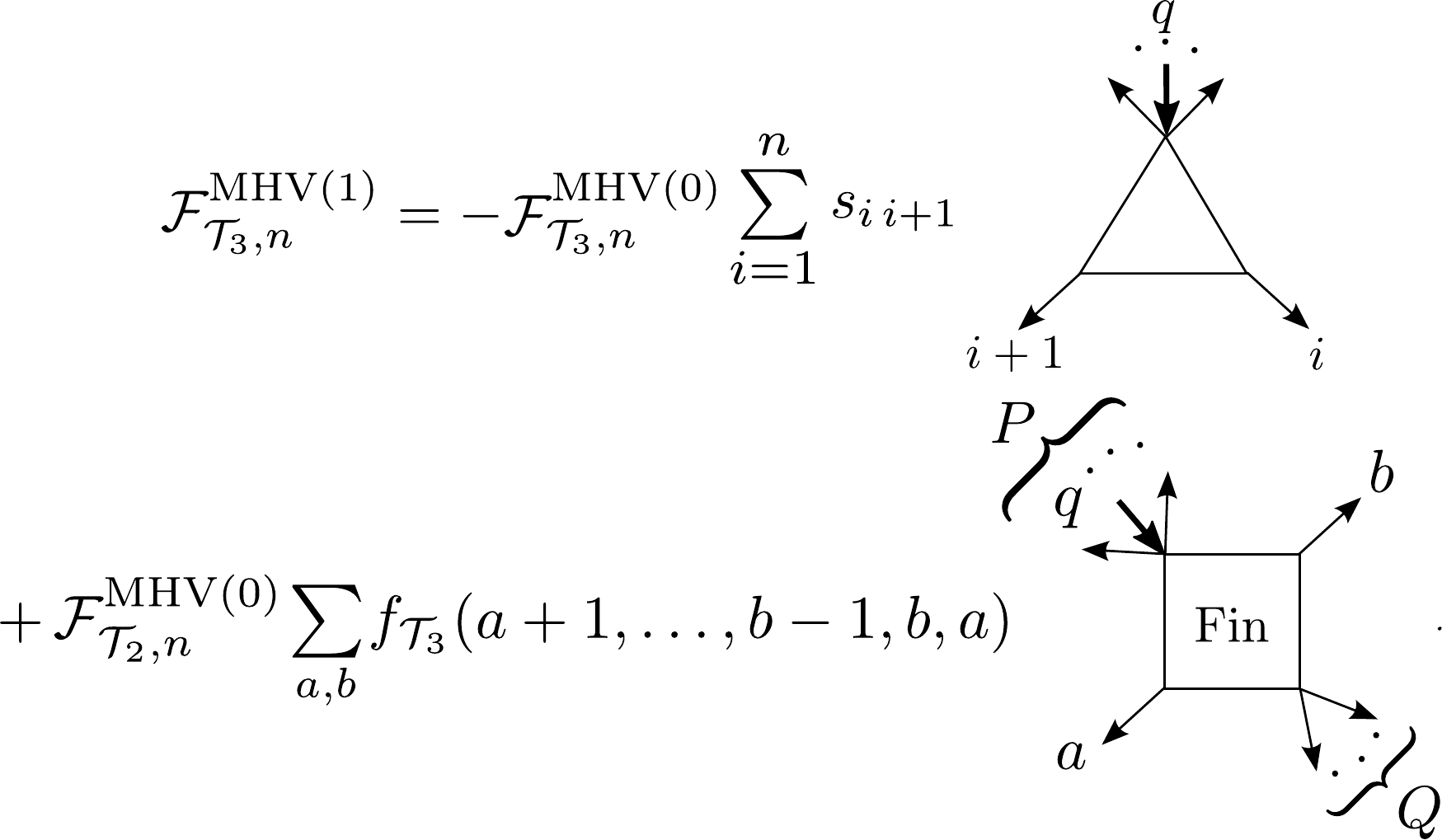}
\caption{\it One-loop result for $\F_{\T_3,n}^{\text{\rm MHV} (1)}$. Here $P$ and $Q$ stand for the momenta of the massive corners and, as usual, $s:=(P+p_a)^2,\;t:=(Q+p_a)^2$ .
}
\label{fig:1loopfullanswer}
\end{figure}

We are now ready to write down the full result for the one-loop MHV super form factor $\F_{\T_3,n}^{\text{MHV} (1)}$ for general $n$. It is given by 
\begin{align}
\begin{split}
\label{eq:1-loop-MHV-BIS}
\F_{\T_3,n}^{\text{MHV} (1)}\ &=  \  \F_{\T_3,n}^{\text{MHV}(0)}\,    \sum_{i=1}^n {  ( - s_{i\,i+1})^{- \epsilon}   \over \epsilon^2}  \\
   &+ \F_{\T_2,n}^{\text{MHV}(0)} \sum_{a, b} f_{\T_3}(a+1,\dots,b-1,b,a)  \text{Fin}^{\rm 2me}(p_a, p_b, P, Q)\, .   
\end{split}
\end{align}
%
For clarity, we illustrate \eqref{eq:1-loop-MHV-BIS} graphically in Figure \ref{fig:1loopfullanswer}.

\subsection{$n$-point MHV super form factors of $\T_k$}

The one-loop result for general $k$ is not qualitatively different from that for $k=3$ computed in the previous section; the only work to do is to calculate the coefficients of the finite two-mass easy boxes using quadruple cuts. Happily, most of the work has already been done. Indeed,  once we know the result for $\F_{\T_3,n}^{\text{MHV}(1)}$, the generalisation for $\F_{\T_k,n}^{\text{MHV}(1)}$ is almost immediate. This is because  the tree-level result  \eqref{conj}  for $\F_{\T_k,n}^{\text{MHV}(0)}$ has  the same trivial dependence on legs $n-1$ and $n$ as $\F_{\T_3,n}^{\text{MHV}(0)}$. The answer is then an immediate generalisation of \eqref{eq:1-loop-MHV-BIS}:
\beqa
\label{eq:1-loop-MHV-all-k}
\begin{split} 
\F_{\T_k,n}^{\text{MHV} (1)} &=   \    
\F_{\T_k,n}^{\text{MHV}(0)}\, 
\sum_{i=1}^n 
{  ( - s_{i\,i+1})^{- \epsilon}   \over \epsilon^2}  \\
   &+  \, \F_{\T_2,n}^{\text{MHV}(0)} \sum_{a, b} f_{\T_k}(a+1,\dots,b-1,b,a)  \ \text{Fin}^{\rm 2me}(p_a, p_b, P, Q)\ .   
\end{split} 
\eeqa
This is our final, compact  expression for the  $n$-point  form factor of $\cT_k$ at one loop with arbitrary $k$ and $n$.


\vspace{0.7cm}

\section*{Acknowledgements}

It is a pleasure to thank  Massimo Bianchi, Andi Brandhuber, Paul Heslop and Yu-tin Huang for discussions.  This work 
was supported by the Science and Technology Facilities Council Consolidated Grant ST/J000469/1  
{\it String theory, gauge theory \& duality. }

\vspace{0.7cm}


\appendix

\section{Scalar integrals}
\label{app:integrals}
\begin{figure}[htb]
\centering
\includegraphics[width=\linewidth]{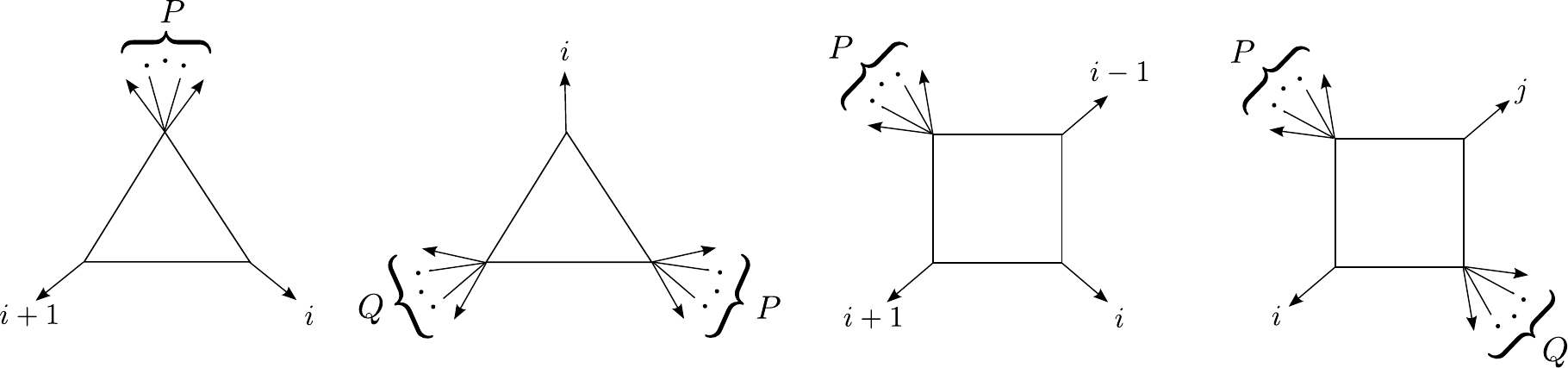}
\caption{\it Scalar integrals which appear in the calculation of the one-loop form factors considered in this paper.}
\label{fig:scalarints}
\end{figure}
In this appendix we give the explicit expressions of various integral functions used in this paper. For the definition of the various momenta we refer the reader to  Figure \ref{fig:scalarints}.\\[12pt]
The one-mass and two-mass triangle integrals are given by
\begin{align}
\label{eq:1massTri}
I_{3;i}^{1m}(P^2)&=\frac{r_\Gamma}{\epsilon ^2} (- P^2)^{-1-\epsilon}\ ,\\
\label{eq:2massTri}
I_{3;i}^{2m}(P^2,Q^2)&=\frac{r_\Gamma}{\epsilon ^2} \frac{(- P^2)^{-\epsilon}-(- Q^2)^{-\epsilon}}{P^2-Q^2} \ , 
\end{align}
where 
\beq
\label{rgamma}
 r_\Gamma \, :=\,  
 \dfrac{\Gamma(1+\epsilon)\Gamma^2(1-\epsilon)}{\Gamma(1-2\epsilon)}
 \ . 
 \eeq 
The two-mass easy box function is given by%
\footnote{The relation between the scalar integral $I$ and the box function $F$  is $I^{\rm 2me}_{4;i,j}= {2r_\Gamma}F^{\rm 2me}_{4;i,j}/ (P^2Q^2-st)$ \cite{Bern:1994zx}.}
\beqa
\label{eq:2me}
F^{\rm 2me}_{4;i,j}(s,t,P^2,Q^2)  
& =  &  
- \frac{1}{\epsilon^2}  \Big[ 
(-s)^{-\epsilon}  +(-t)^{-\epsilon}-(-P^2)^{-\epsilon}-(-Q^2)^{-\epsilon} \Big]  \nonumber \\ \cr
&+ & {\rm Fin^{2me}}
(s,t,P^2,Q^2) 
\ ,
\eeqa
where 
 $s:= (p_i+P)^2$ and  $t:= (p_i+Q)^2$.
The  finite part of the  two-mass easy box function, in the form of  \cite{Duplancic:2000sk,Brandhuber:2004yw}, is 
\beq
\label{2mebst}
  {\rm  Fin^{2me}} (s,t,P^2, Q^2) \, = \, 
 \Li(1-aP^2)\, + \, \Li(1-aQ^2)  \, -\,  \Li(1-as)
\,  -\,  \Li(1-at)
\ , 
\eeq
where 
\beq
\label{adef} 
a \ = \
\frac{P^2+Q^2-s-t}{P^2Q^2-st} \ . 
\eeq
An analytic proof of the equivalence of \eqref{2mebst} and the form given in 
\cite{Bern:1993kr}  can be found in \cite{Brandhuber:2004yw}. 
Finally, we note that  one-mass box functions can be obtained from the two-mass easy ones by simply taking the limit $P^2 \to0$ \cite{Bern:1993kr}.

\section{Cyclicity of $\F^{\rm MHV}_{\T_4,n}$} \label{app:cyclicity}

In this appendix we prove the cyclicity of the form factor $\F^{\rm MHV}_{\T_4,n}$. This is given in \eqref{eq:O4}, but for convenience  we repeat its expression here:
\begin{align}
\F^{\text{MHV}}_{\T_4,n}\ = \ \F^{\text{MHV}}_{\T_2,n} \sum\limits_{1\leq i\leq j}^{n-3} \sum\limits_{j< k\leq l}^{n-2} (2-\delta_{ij})(2-\delta_{kl}) \frac{\b{n\,i}\b{j\,k}\b{l\,n-1}}{\b{n-1\,n}}(\eta_{-, i}\cdot \eta_{-,j})(\eta_{-,k}\cdot \eta_{-,l})\, .
\end{align}
The procedure we will follow consists in  eliminating $\eta_{-, 1}$ using supermomentum conservation in the $Q_{-}$ direction,  
and showing that the result one obtains in this way is the same as the original expression but with all relevant indices shifted by one unit. 
After substituting in the solution for $\eta_{-,1}$ from supermomentum conservation,  
we consider contributions to  terms of different structure in the various  $\eta_{-}$'s separately. 
In what follows we will list all possible structures and their corresponding coefficients:

\begin{itemize}

\item $(\eta_{-,i} \cdot \eta_{-,j})(\eta_{-,n-1})^2$ : 
\begin{align}
(2-\delta_{ij})\frac{\b{n\,1}\b{1\,i}\b{j\,n\!-\!1}}{\b{n\!-\!1\,n}} \frac{\b{n\!-\!1 \, n}^2 }{\b{n \, 1 }^2} \ = \  (2-\delta_{ij}) \frac{\b{1\, i}\b{j\,n\!-\!1}\b{n\!-\!1\,n}}{\b{n\,1}} \, .
\end{align}

\item $(\eta_{-,i})^2(\eta_{-,k})^2, \quad {\rm with} \quad i<k$: 
\begin{align}
\begin{split}
&\frac{\b{n\,i}\b{i\,k}\b{k\,n\!-\!1}}{\b{n\!-\!1\,n}}
+
\frac{\b{n\,1}\b{1\,i}\b{i\,n\!-\!1}}{\b{n\!-\!1\,n}}\frac{\b{k \, n}^2 }{\b{n \, 1}^2}
+
\frac{\b{n\,1}\b{1\,k}\b{k\,n\!-\!1}}{\b{n\!-\!1\,n}}\frac{\b{i \, n}^2 }{\b{n \, 1}^2}\\
-&2 \frac{\b{i\,n}\b{k\,n}}{\b{n \, 1}^2} \frac{\b{n\,1}\b{1\,i}\b{k\,n\!-\!1}}{\b{n\!-\!1 \, n}}
+
2\frac{\b{i\,n}}{\b{n \, 1}} \frac{\b{n\,1}\b{i\,k}\b{k\,n\!-\!1}}{\b{n\!-\!1 \, n}}
 \ = \  \frac{\b{1 \, i} \b{i \, k} \b{k \, n } }{\b{n \, 1 }}\, .
\end{split}
\end{align}

\item $(\eta_{-,i} \cdot \eta_{-,j})(\eta_{-,k})^2, \quad {\rm with} \quad i< j<k$ :
\begin{align}
\begin{split}
&2\frac{\b{n\,i}\b{j\,k}\b{k\,n\!-\!1}}{\b{n\!-\!1\,n}}
+
2\frac{\b{n\,1}\b{1\,i}\b{j\,n\!-\!1}}{\b{n\!-\!1\,n}}\frac{\b{k \, n}^2 }{\b{n \, 1}^2}
+
2\frac{\b{n\,1}\b{1\,k}\b{k\,n\!-\!1}}{\b{n\!-\!1\,n}}\frac{\b{i \, n} \b{j\, n} }{\b{n \, 1}^2}\\
-&2 \frac{\b{j\,n}\b{k\,n}}{\b{n \, 1}^2} \frac{\b{n\,1}\b{1\,i}\b{k\,n\!-\!1}}{\b{n\!-\!1 \, n}}
-2 \frac{\b{i\,n}\b{k\,n}}{\b{n \, 1}^2} \frac{\b{n\,1}\b{1\,j}\b{k\,n\!-\!1}}{\b{n\!-\!1 \, n}}\\
-&
2\frac{\b{k\,n}}{\b{n \, 1}} \frac{\b{n\,1}\b{i\,j}\b{k\,n\!-\!1}}{\b{n\!-\!1 \, n}} 
+
2\frac{\b{j\,n}}{\b{n \, 1}} \frac{\b{n\,1}\b{i\,k}\b{k\,n\!-\!1}}{\b{n\!-\!1 \, n}}
+
2\frac{\b{i\,n}}{\b{n \, 1}} \frac{\b{n\,1}\b{j\,k}\b{k\,n\!-\!1}}{\b{n\!-\!1 \, n}}\\
 \ = \  &2 \frac{\b{1 \, i} \b{j \, k} \b{k \, n } }{\b{n \, 1 }} \, .
\end{split}
\end{align}

\item $(\eta_{-,i} \cdot \eta_{-,j})(\eta_{-,k})^2, \quad {\rm with} \quad k<i< j $ :
\begin{align}
\begin{split}
&2\frac{\b{n\,k}\b{k\,i}\b{j\,n\!-\!1}}{\b{n\!-\!1\,n}}
+
2\frac{\b{n\,1}\b{1\,i}\b{j\,n\!-\!1}}{\b{n\!-\!1\,n}}\frac{\b{k \, n}^2 }{\b{n \, 1}^2}
+
2\frac{\b{n\,1}\b{1\,k}\b{k\,n\!-\!1}}{\b{n\!-\!1\,n}}\frac{\b{i \, n} \b{j\, n} }{\b{n \, 1}^2}\\
-&2 \frac{\b{k\,n}\b{j\,n}}{\b{n \, 1}^2} \frac{\b{n\,1}\b{1\,k}\b{i\,n\!-\!1}}{\b{n\!-\!1 \, n}}
-2 \frac{\b{k\,n}\b{i\,n}}{\b{n \, 1}^2} \frac{\b{n\,1}\b{1\,k}\b{j\,n\!-\!1}}{\b{n\!-\!1 \, n}}
\\
+&
4\frac{\b{k\,n}}{\b{n \, 1}} \frac{\b{n\,1}\b{k\,i}\b{j\,n\!-\!1}}{\b{n\!-\!1 \, n}} 
 \ = \ 2 \frac{\b{1 \, k} \b{k \, i} \b{j \, n } }{\b{n \, 1 }} \, .
\end{split}
\end{align}

\item $(\eta_{-,i} \cdot \eta_{-,j})(\eta_{-,k})^2, \quad {\rm with} \quad i<k< j$ : 
\begin{align}
\begin{split}
&
2\frac{\b{n\,1}\b{1\,i}\b{j\,n\!-\!1}}{\b{n\!-\!1\,n}}\frac{\b{k \, n}^2 }{\b{n \, 1}^2}
+
2\frac{\b{n\,1}\b{1\,k}\b{k\,n\!-\!1}}{\b{n\!-\!1\,n}}\frac{\b{i \, n} \b{j\, n} }{\b{n \, 1}^2}\\
-&2 \frac{\b{k\,n}\b{j\,n}}{\b{n \, 1}^2} \frac{\b{n\,1}\b{1\,i}\b{k\,n\!-\!1}}{\b{n\!-\!1 \, n}}
-2 \frac{\b{k\,n}\b{i\,n}}{\b{n \, 1}^2} \frac{\b{n\,1}\b{1\,k}\b{j\,n\!-\!1}}{\b{n\!-\!1 \, n}}
\\
- &
2\frac{\b{k\,n}}{\b{n \, 1}} \frac{\b{n\,1}\b{i\,k}\b{j\,n\!-\!1}}{\b{n\!-\!1 \, n}}+
2\frac{\b{j\,n}}{\b{n \, 1}} \frac{\b{n\,1}\b{i\,k}\b{k\,n\!-\!1}}{\b{n\!-\!1 \, n}} 
 \ = \ 0  \, .
\end{split}
\end{align}

\item $(\eta_{-,i} \cdot \eta_{-,j})(\eta_{-,k} \cdot \eta_{-,l}), \quad {\rm with} \quad i<j<k<l$ :
\begin{align}
\begin{split}
&
4\frac{\b{n\,i}\b{j\,k}\b{l\,n\!-\!1}}{\b{n\!-\!1\,n}}
+
4\frac{\b{n\,1}\b{1\,i}\b{j\,n\!-\!1}}{\b{n\!-\!1\,n}}\frac{\b{k \, n} \b{l\, n} }{\b{n \, 1}^2}
+
4\frac{\b{n\,1}\b{1\,k}\b{l\,n\!-\!1}}{\b{n\!-\!1\,n}}\frac{\b{i \, n} \b{j\, n} }{\b{n \, 1}^2}\\
+& 4 \frac{\b{j\,n}}{\b{n \, 1}} \frac{\b{n\,1}\b{i\,k}\b{l\,n\!-\!1}}{\b{n\!-\!1 \, n}}
+4 \frac{\b{i\,n}}{\b{n \, 1}} \frac{\b{n\,1}\b{j\,k}\b{l\,n\!-\!1}}{\b{n\!-\!1 \, n}} 
-4 \Big[ \frac{\b{k\,n}}{\b{n \, 1}} \frac{\b{n\,1}\b{i\,j}\b{l\,n\!-\!1}}{\b{n\!-\!1 \, n}} \\
+&
\frac{\b{i\,n}\b{k\,n}}{\b{n \, 1}^2 } \frac{\b{n\,1}\b{1\,j}\b{l\,n\!-\!1}}{\b{n\!-\!1 \, n}} 
+
\frac{\b{j\,n}\b{l\,n}}{\b{n \, 1}^2 } \frac{\b{n\,1}\b{1\,i}\b{k\,n\!-\!1}}{\b{n\!-\!1 \, n}} \Big]
 \ = \  
 4\frac{\b{1\,i}\b{j\,k}\b{l\,n}}{\b{n \, 1}} \, .
\end{split}
\end{align}

\item $(\eta_{-,i} \cdot \eta_{-,j})(\eta_{-,k} \cdot \eta_{-,l}), \quad {\rm with} \quad i<k<j<l$ :
\begin{align}
\begin{split}
&
4\frac{\b{n\,1}\b{1\,i}\b{j\,n\!-\!1}}{\b{n\!-\!1\,n}}\frac{\b{k \, n} \b{l\, n} }{\b{n \, 1}^2}
+
4\frac{\b{n\,1}\b{1\,k}\b{l\,n\!-\!1}}{\b{n\!-\!1\,n}}\frac{\b{i \, n} \b{j\, n} }{\b{n \, 1}^2}\\
+ & 4 \frac{\b{j\,n}}{\b{n \, 1}} \frac{\b{n\,1}\b{i\,k}\b{l\,n\!-\!1}}{\b{n\!-\!1 \, n}} 
-4 \Big[ \frac{\b{l\,n}}{\b{n \, 1}} \frac{\b{n\,1}\b{i\,k}\b{j\,n\!-\!1}}{\b{n\!-\!1 \, n}} \\
+ &
\frac{\b{k\,n}\b{j\,n}}{\b{n \, 1}^2 } \frac{\b{n\,1}\b{1\,i}\b{l\,n\!-\!1}}{\b{n\!-\!1 \, n}} 
+
\frac{\b{i\,n}\b{l\,n}}{\b{n \, 1}^2 } \frac{\b{n\,1}\b{1\,k}\b{j\,n\!-\!1}}{\b{n\!-\!1 \, n}} \Big]
 \ = \ 
0\, .
\end{split}
\end{align}

\end{itemize}
Thus we have shown that all terms $(\eta_{-,i} \cdot \eta_{-,j})(\eta_{-,k} \cdot \eta_{-,l})$ with the right ordering, namely when $i \leq j < k \leq l$, have the correct coefficients, whereas when $i,j,k,l$ are in a wrong ordering the corresponding coefficients vanish. This completes the proof of the cyclicity of $\F_{\T_4,n}$. 


\bibliographystyle{utphys}
\bibliography{bibliography}


\end{document}